\documentclass[prb,aps,onecolumn,superscriptaddress]{revtex4-2}
\usepackage{graphicx}
\usepackage{booktabs} 
\usepackage{dcolumn}
\usepackage{bm}
\usepackage{comment} 
\usepackage{lipsum}
\usepackage{ragged2e}
\usepackage{braket}
\usepackage{bbm}
\usepackage{lipsum}
\usepackage{color}
\usepackage{xcolor}
\usepackage{amsfonts}
\usepackage{amsmath}
\usepackage{amsmath,amssymb,latexsym}
\usepackage{wasysym,stmaryrd,marvosym}
\usepackage{mathrsfs}
\usepackage{dsfont}
\usepackage{float}
\providecommand{\keywords}[1]{\textbf{\textit{Index terms---}} #1}
\usepackage[mathlines]{lineno}
\usepackage[colorinlistoftodos]{todonotes}
\usepackage[colorlinks=true, allcolors=blue]{hyperref}
\usepackage{soul}
\newcommand{\MA}[1]{{\color{purple}{#1}}} 
\newcommand{\JM}[1]{{\color{red}{#1}}} 

\begin{document}
\newcommand{\cnyn}{Centro de Nanociencias y Nanotecnolog\'ia,
Universidad Nacional Aut\'onoma de M\'exico, Apartado Postal 2681, 22800, Ensenada, Baja California, M\'exico.}
\newcommand{\uabc}{Facultad de Ciencias, Universidad Aut\'onoma de Baja California, Apartado Postal 1880, 22800 Ensenada, Baja California, M\'exico}
\newcommand{\ohio}{Department of Physics and Astronomy, Ohio University, Athens, Ohio 45701, USA}
\newcommand{\ubc}{Department of Physics and Astronomy and Quantum Matter Institute,
University of British Columbia, Vancouver, British Columbia BC V6T 1Z4, Canada}

\title{
Optical spin Edelstein response of $\boldsymbol{d}$-wave altermagnets}

\author{M. A. Mojarro}
\email{miguel.mojarroramirez@ubc.ca}
\affiliation{\ubc}
\author{R. Carrillo-Bastos}
\email{ramoncarrillo@uabc.edu.mx}
\affiliation{\ohio}
\affiliation{\uabc}
\author{Jes\'us A. Maytorena}
\email{jesusm@ens.cnyn.unam.mx}
\affiliation{\cnyn}

\date{\today}

\begin{abstract}
The momentum-dependent spin splitting and compensated magnetic order in altermagnets offer opportunities for electrical control of spin polarization. 
We explore the spin Edelstein effect in the frequency domain of
a two-dimensional $d$-wave altermagnet with Rashba spin-orbit coupling.
We calculate the spin-velocity
susceptibility response function beyond the intraband response
regime 
and find that the interband longitudinal and transverse Edelstein responses 
are governed by the Zeeman quantum metric and Zeeman Berry curvature, respectively.
The density of states presents a logarithmic van Hove singularity
at exactly the negative value of the energy below which four disconnected Fermi contours emerge. This defines three distinctive regions for the position of the Fermi level,
through which the Edelstein susceptibility tensor displays characteristic spectral features,
reflecting the topological change of the Fermi surface.
The resulting
spin magnetization shows a strong dependence on the driving frequency and on 
the orientation of the in-plane external electric field. Our results
highlight new possibilities for electrical and optical control of the spin Edelstein effect in altermagnets that could find spintronic applications.

\end{abstract}


\maketitle

\section{Introduction}

Despite the absence of net magnetization, altermagnets exhibit momentum-dependent spin splitting in their energy band structure of nonrelativistic origin  \cite{Libor2022_1,Libor2022_2}. This splitting displays $d$-, $g$-, or $i$-wave symmetry, depending on the underlying crystal structure and magnetic order \cite{Libor2022_1,Jungwirth2026_2}. Recent angle-resolved photoemission spectroscopy measurements have provided evidence of altermagnetic spin splitting in several candidate materials \cite{Krempasky2024,Ding2024,Jiang2025}. These properties make altermagnets promising platforms for exploring unconventional electronic and spin-transport phenomena \cite{Bai2023,Bai2024,Jungwirth2026}. Examples include anomalous Hall effects in the absence of external magnetic fields \cite{Gonzalez2023,Reichlova2024}, strain-dependent Hall responses \cite{Liu2026}, nonlinear Hall effects \cite{Chu2025}, and anomalous Nernst effects \cite{Badura2025}.  Beyond charge transport, altermagnets also support unconventional spin-transport phenomena, including spin-polarized currents \cite{GonzalezHernandez2021SpinSplitter} and current-induced spin torques \cite{Vakili2025SpinTorque}.
In particular, the possibility of controlling the spin transport properties of altermagnets through external electric and electromagnetic fields offers an interesting direction for altermagnetic spintronics.

The spin Edelstein effect (SEE) provides a mechanism for the electrical control of spin, in which an external electric field induces a spin polarization in systems with broken spatial inversion symmetry \cite{Annika2016,TheorySEE}.
The theoretical study of the SEE has mainly focused on the intraband
response regime in the dc limit 
\cite{TheorySEE,Annika2016,SpinPolEng,Cao2025,Golub2025,Iurov2026,Nonlinear-anomalous,Slow-phonon-control} or
on driving frequencies below the onset of interband transitions \cite{LXiangPRLb,DESG}.
However, the finite frequency response regime is interesting in itself because
it extends the phenomena to coupling with electromagnetic waves and optical probing.
The frequency dependence of the current-induced spin polarization
in spin-orbit coupled quantum wells formed in semiconductor heterostructures 
reveals notable features such as a strong dispersion of the spin polarization in the vicinity
of a fixed precession axis \cite{Raichev} and an anisotropic response induced by a
nonisotropic momentum space available for interband transitions caused by anisotropic spin-splitting \cite{Priscilla2010}. In noncentrosymmetric antiferromagnets CuMnAs and Mn$_2$Au, density-functional calculations show sizable frequency-dependent induced spin and orbital Rashba Edelstein magnetizations at optical frequencies \cite{Salemi2019}. 

In altermagnets, the coexistence of nonrelativistic spin splitting and spin-orbit coupling offers additional possibilities for controlling the spin polarization via external electric fields \cite{SpinPolEng,Cao2025,Golub2025,Slow-phonon-control,Iurov2026}.
For instance, engineering of the spin and orbital Edelstein effects is suggested
in $d$-wave altermagnets via Rashba spin-orbit interaction (RSOI), controlled
by electrostatic gating, and through the application of off-resonant illumination \cite{SpinPolEng}.
In such a case, a highly controllable dc spin polarization, with in-plane and out-of-plane components, is reported in the experimentally observable range.
Nonrelativistic SEE has also been investigated in unconventional $p$-wave magnets, demonstrating that the spin control via electric fields can be achieved without spin-orbit interaction in related altermagnetic systems \cite{Chakraborty2025}. Nevertheless, the frequency-dependent SEE in altermagnets, particularly its interband contribution, remains an important aspect to explore.

Here, 
we investigate the spin polarization
induced by an ac electric field in a 2D $d$-wave altermagnet in the presence of
RSOI. Within the Kubo formalism, we consider the intrinsic SEE and calculate
the frequency-dependent spin-velocity susceptibility tensor over the full range of Fermi energies, including the negative energy part of the band structure, as this region 
has been recognized to exhibit novel and interesting features such as tunable superconducting phases \cite{Supercond_RashbaGases} and large Hall angles \cite{Chen_2025}.


To address the role played by quantum geometry, we explore the dynamical spin Edelstein
susceptibility within the framework of the recently introduced Zeeman quantum geometric tensor
\cite{LXiangPRLa,LXiangPRLb}. This quantum object has been shown to play a key role in generating a gyrotropic magnetic current \cite{LXiangPRLa} and identifying nonrelativistic spin-splitting symmetries \cite{ZQGT-SciPost}. 
Here, we show how it is involved
in determining the interband part of the optical (intrinsic) SEE of the altermagnet under consideration. We find that the interband longitudinal Edelstein susceptibility is connected
to the diagonal elements of the Zeeman quantum metric, while the off-diagonal
elements of the Zeeman Berry curvature are responsible for the transverse 
spin polarization. Moreover, due to the symmetry of the Zeeman quantum geometric tensor
of the present model, it is possible to classify the real and imaginary parts of the susceptibility components into both reactive-type and dissipative-type response functions, 
in analogy to the causal optical conductivity response. These functions display characteristic
spectral features, which evolve distinctly as the Fermi level varies and which
serve as optical fingerprints of the altermagnetic order.
Finally, we show that the magnitude and orientation of the induced in-plane spin polarization can be manipulated by varying the driving frequency and the orientation of the applied electric field, providing a dynamical route for controlling spin polarization in $d$-wave altermagnets.

The paper is organized as follows. In Section \ref{Sec2}, we introduce the Hamiltonian model
and study ground state properties such as energy dispersion, spin texture, Fermi contours,
and density of states. Section \ref{Sec3} presents the
Kubo formula for the dynamical susceptibility response function characterizing the
SEE in the frequency domain. The interband contribution is expressed in
terms of the Zeeman quantum geometric tensor, and explicit expressions for a 
generic two-band model are derived. The calculated susceptibility spectra are analyzed.
Section \ref{Sec4} contains a discussion on the spin polarization vector field and its
dependence on the driving frequency and the orientation of the in-plane electric field.
Section \ref{Sec5} is devoted to conclusions.

\section{Rashba coupled $\boldsymbol{d}$-wave altermagnet}\label{Sec2}

We consider a 2D $d$-wave altermagnet in the presence of Rashba spin-orbit interaction (RSOI),
with a momentum-space Hamiltonian \cite{SpinPolEng,Rao2024,Slow-phonon-control} 
\begin{equation} \label{Ham}
\hat{H}({\bf k})=\frac{\hbar^2(k_x^2+k_y^2)}{2m}\hat{\sigma}_0 + \alpha(k_y{\bf\hat{x}}-k_x{\bf\hat{y}})\cdot\boldsymbol{\hat{\sigma}}+h({\bf k})\hat{\sigma}_z\,,
\end{equation}
where $\hat{\sigma}_0$ is the $2\times 2$ identity matrix, $\boldsymbol{\hat{\sigma}}$
is the vector of Pauli matrices for the spin degree of freedom,
$h({\bf k})={\cal D}\hbar^2(k_x^2-k_y^2)/2m$; $\alpha$ and ${\cal D}\leqslant 1$
are the RSOI and the $d$-wave strength parameters, respectively; we shall take ${\cal D}<1$ in this paper, as this guarantees closed Fermi contours. 
The Hamiltonian (\ref{Ham})
lacks spatial inversion ($\mathcal{P}$) and time-reversal ($\mathcal{T}$) symmetries due to RSOI and $h({\bf k})$, respectively, but preserves the composite symmetry operation $C_{4z}\hat{\mathcal{T}}$, where $C_{4z}$ is a four-fold rotation around the $z$-axis and
$\hat{\mathcal{T}}=-i\hat{\sigma}_y\hat{\mathcal{K}}$ is the time-reversal operator with $\hat{\mathcal{K}}$ the complex conjugate operator.
We can verify that $(C_4\hat{\mathcal{T}})\hat{H}({\bf k})(C_4\hat{\mathcal{T}})^{-1}=\hat{H}(C_4\hat{\mathcal{T}}{\bf k})=\hat{H}(k_y,-k_x)$ \cite{Gonzalez2025}.
We note that this model can be realized in the vicinity of the $\Gamma$ point of a $d$-wave altermagnet with inversion symmetry breaking.

\begin{figure}
\includegraphics[scale=0.32]{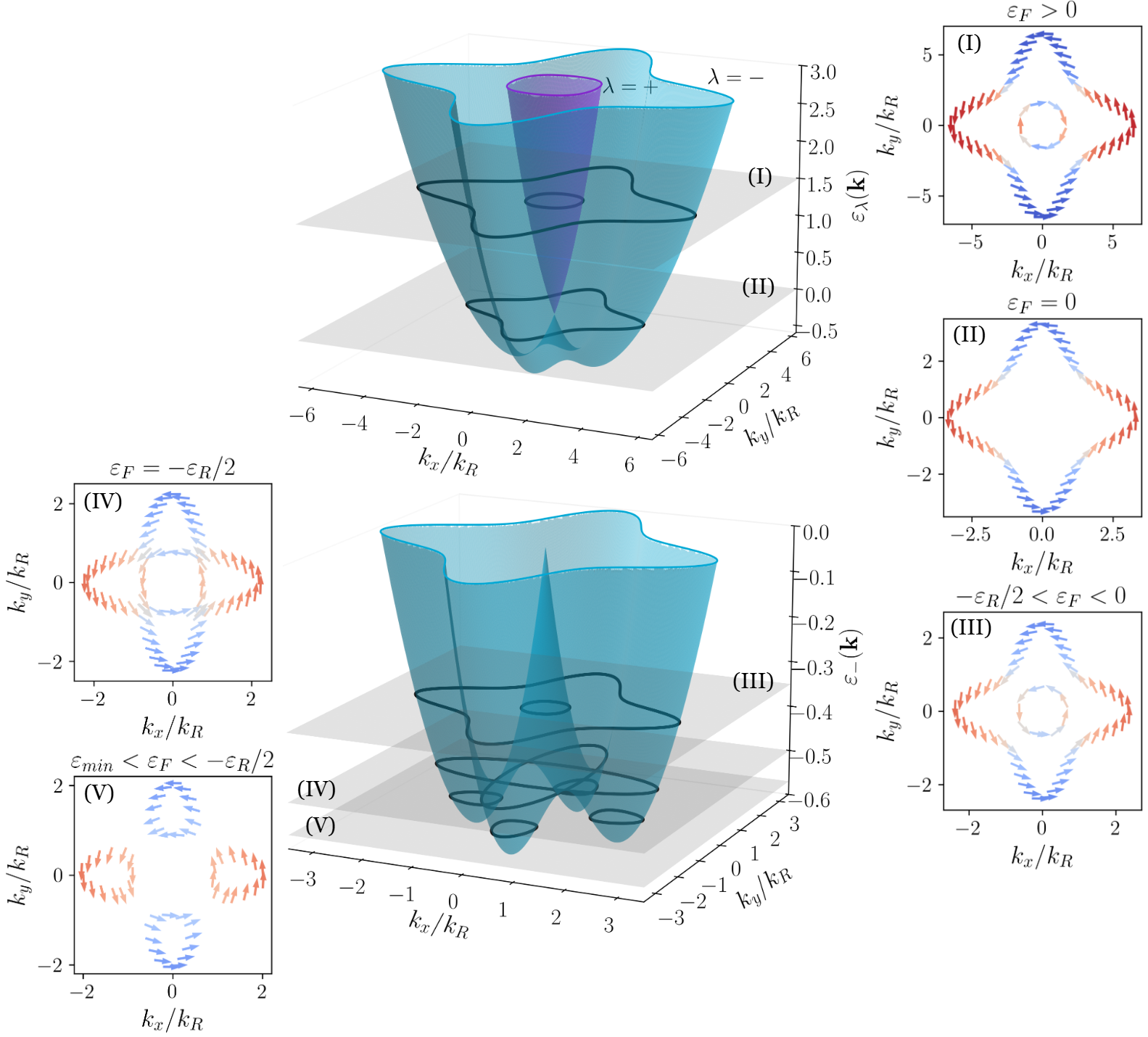}
\caption{Band structure of a 2D Rashba $d$-wave altermagnet (normalized to $
\varepsilon_R=\alpha k_R=m\alpha^2/\hbar^2$) and the topological change of the Fermi surface as Fermi energy changes. The energy $\varepsilon=0$ is set to the degeneracy point ${\bf k}={\bf 0}$.
The lower surface depicts the negative energy part of
the band structure shown in the upper figure. Each inset corresponds to a cut of the band
structure at a given value of the Fermi level $\varepsilon_F$ (light gray planes: cuts
I and II for $\varepsilon_F\geqslant0$, cuts III to V for $\varepsilon_F<0$).
Note how the Fermi contours (black solid lines) evolve from cut to cut.
The arrows indicate the in-plane spin orientation, while the color indicates the out-of-plane orientation, with red (blue) corresponding to spin-up (spin-down).
}
\label{Surf}
\end{figure}

The corresponding energy-momentum dispersion is
\begin{equation} \label{dispersions}
\varepsilon_{\lambda}(k_x,k_y)=\frac{\hbar^2(k_x^2+k_y^2)}{2m}+\lambda\, d(k_x,k_y)\,,
\end{equation}
where $\lambda=\pm$ defines the helicity of the states in the upper ($+$) and lower ($-$) bands
(Fig.\,\ref{Surf}), and
\begin{equation} \label{dmagnitud}
d(k_x,k_y)=\sqrt{\alpha^2(k_x^2+k_y^2)+(\hbar^2{\cal D}/2m)^2(k_x^2-k_y^2)^2}\,.
\end{equation}
This function determines the energy difference between the bands as
$\varepsilon_+({\bf k})-\varepsilon_-({\bf k})=2d({\bf k})$. We also note that only the $\lambda=-$ band takes negative values (Fig.\,\ref{Surf}). 
 
To illustrate the rotational $C_{4z}$-symmetry introduced by the quadratic altermagnetic energy
$h({\bf k})$ (Fig.\,\ref{Surf}), 
we rewrite the energy (\ref{dmagnitud}) in polar coordinates for
the wave vector ${\bf k}=k_x{\bf\hat{x}}+k_y{\bf\hat{y}}=k(\cos\theta{\bf\hat{x}}+
\sin\theta{\bf\hat{y}})$,
\begin{equation} \label{ladé}
d(k,\theta)=\frac{\hbar^2}{2m}k\sqrt{(2k_R)^2+k^2{\cal D}^2\cos^22\theta}\,.
\end{equation}
 
For a given direction $\theta\in[0,2\pi]$,
$-\varepsilon_R/(1+\sqrt{1-{\cal D}^2\cos^22\theta})<\varepsilon_-(k,\theta)<0$ for 
$k\in[0,q_0(\theta)]$, where $k_R=m\alpha/\hbar^2$ and 
$\varepsilon_R=\alpha k_R$ are the characteristic momentum and energy associated with 
the RSOI, respectively, and $q_0(\theta)=2k_R/\sqrt{1-{\cal D}^2\cos^22\theta}$.
The absolute minimum value of the band $\varepsilon_-$
is $\varepsilon_{min}$, with
$\varepsilon_{min}=-\varepsilon_R/(1+\sqrt{1-{\cal D}^2})$, reached at angles for which $\cos^22\theta=1$.
We also note that for $\alpha=0$, the energy bands are degenerate along the lines $k_x=\pm k_y$, whereas this degeneracy is lifted for $\alpha\neq0$ at finite momenta, persisting only at $k=0$ (Fig.\,\ref{Surf}).

The orientation of the spin in each band is given by the vector field
$\langle\boldsymbol{\sigma}\rangle_{\lambda}=
\langle\lambda|\boldsymbol{\sigma}|\lambda\rangle$,
\begin{equation} \label{spintexture}
\langle\boldsymbol{\sigma}\rangle_{\lambda} =\lambda\,\frac{-2k_R\boldsymbol{\hat{\theta}}_k+{\cal D}k\cos2\theta{\bf\hat{z}}}{\sqrt{(2k_R)^2+{\cal D}^2k^2\cos^22\theta}},
\end{equation}
where $\boldsymbol{\hat{\theta}}_k=(-k_y{\bf\hat{x}}+k_x{\bf\hat{y}})/k=-\sin\theta{\bf\hat{x}}+
\cos\theta{\bf\hat{y}}$ is the tangential unit vector in the $k_xk_y$-plane.
The spin textures (\ref{spintexture}) along the Fermi lines defined by the equation $\varepsilon_{\lambda}(k_x,k_y)=\varepsilon_F$, are shown in Fig.\,\ref{Surf} for several values of the Fermi energy $\varepsilon_F$.
We note that tuning the Fermi level drives topological Lifshitz transitions in the Fermi contours, and
three energy intervals can be identified:

\underline{Interval I: $\varepsilon_F>0$}

In this case, both bands $\varepsilon_+$ and $\varepsilon_-$ are occupied, and there are two
Fermi lines, given in polar coordinates by the parametric equations
\begin{equation} \label{FermiLine_R+A}
k_F^{\pm}(\theta)=\left(\frac{2m/\hbar^2}{1-{\cal D}^2\cos^22\theta}\right)^{1/2}\left(\varepsilon_R+\varepsilon_F\mp
\sqrt{\varepsilon_R^2+2\varepsilon_F\varepsilon_R+
(\varepsilon_F{\cal D}\cos 2\theta)^2}\right)^{1/2}\,, \ \ \ \theta\in[0,2\pi]\,.
\end{equation}

\underline{Interval II: $-\varepsilon_R/2<\varepsilon_F<0$}

Only the lower band $\varepsilon_-$ is occupied. The Fermi lines are given by
the same equations (\ref{FermiLine_R+A}) but with the substitution 
$\varepsilon_F\to-|\varepsilon_F|$ performed,
\begin{equation} \label{FermiLines2}
k_{F,\pm}^-(\theta)= \left(\frac{2m/\hbar^2}{1-{\cal D}^2\cos^22\theta}\right)^{1/2}\left(\varepsilon_R-|\varepsilon_F|\pm\sqrt{\varepsilon_R^2-2|\varepsilon_F|\varepsilon_R+
(|\varepsilon_F|{\cal D}\cos 2\theta)^2}\right)^{1/2}\,, \ \ \ \theta\in[0,2\pi]\,.
\end{equation}
When $\varepsilon_F=0$, there is a single Fermi contour defined by
$k_{F,+}^-(\theta)=q_0(\theta)$.

\begin{figure}
    \centering
    \includegraphics[scale=0.6]{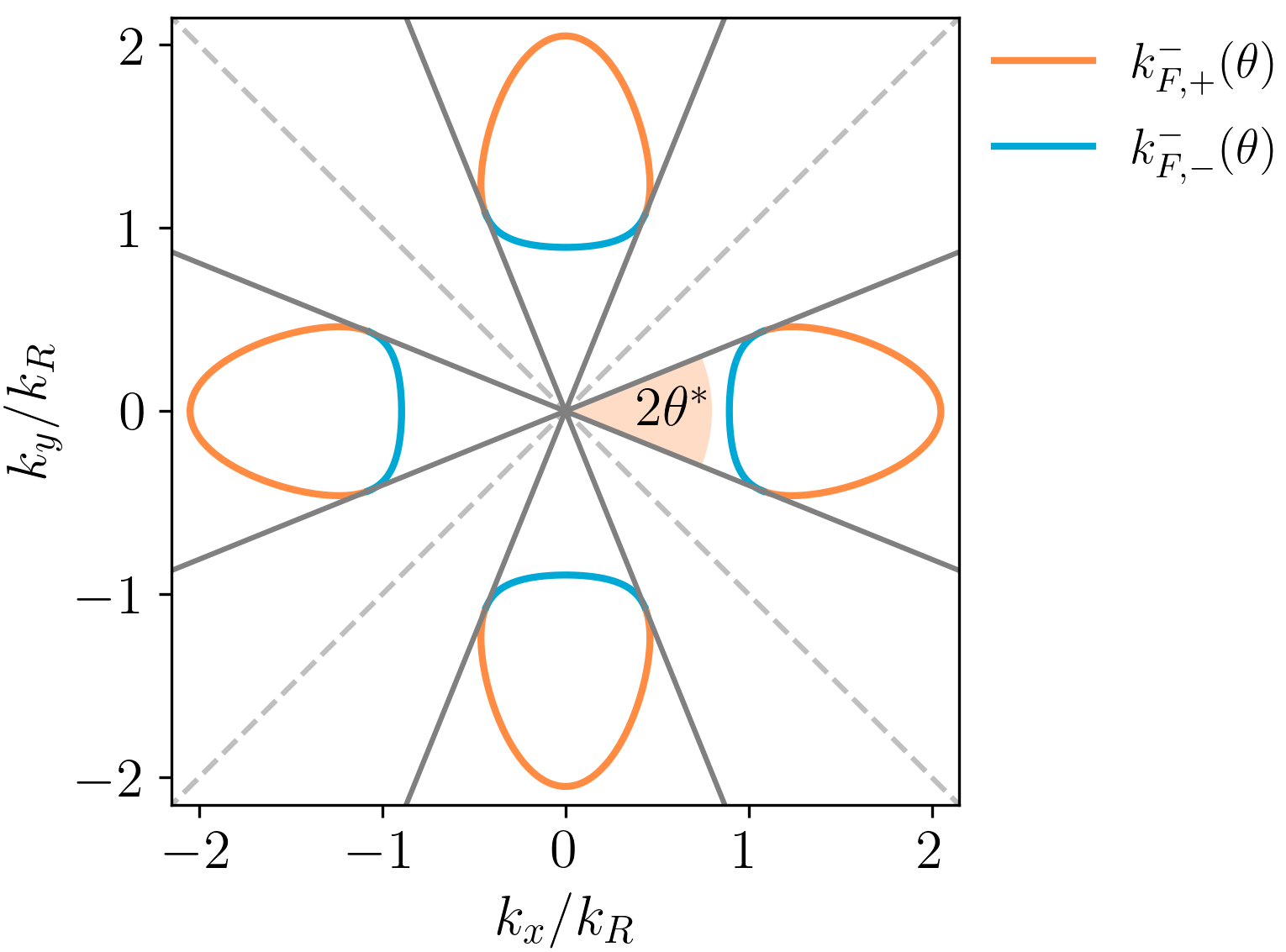}
    \caption{Fermi lines for $\varepsilon_{min}<\varepsilon_F<-\varepsilon_R/2$.
    The sectors with angle $2\theta^*$ between the gray solid lines (centered at $\theta=0, \pi/2, \pi, 3\pi/2$) are defined by the condition (\ref{abscos}). At $\varepsilon_F=\varepsilon_{min}$ the set degenerates into four
    points, while as ${\cal D}\to 0$, the set transforms into a circle of radius 
    $k_R$ at $\varepsilon_F=-\varepsilon_R/2$.}
    \label{disjoint}
\end{figure}

\underline{Interval III: $\varepsilon_{min}<\varepsilon_F<-\varepsilon_R/2$}

Note that at $\varepsilon_F=-\varepsilon_R/2$ these curves reduce to
$k_{F,\pm}^-(\theta)=k_R[(1\pm{\cal D}|\cos2\theta|)/(1-{\cal D}^2\cos^22\theta)]^{1/2}$,
which touch each other when $\cos2\theta=0$ (see Fig.~\ref{Surf}). Below $\varepsilon_F=-\varepsilon_R/2$,
the Fermi contours break into four non-overlapping closed curves (see Fig.~\ref{disjoint}), 
which are still given as for region II, but with the angle 
restricted by the condition
\begin{equation} \label{abscos}
|\cos 2\theta|\geqslant\sqrt{\frac{2\varepsilon_R}{{\cal D}^2\varepsilon_F^2}\left(|\varepsilon_F|-\frac{\varepsilon_R}{2}\right)}\,,
\end{equation}
such that each of the four regions $|k_x|>|k_y|$ and $|k_x|<|k_y|$ (separated by
dashed light gray lines in Fig.~\ref{disjoint})
contains only one of the four closed contours.
 When $\varepsilon_F\to \varepsilon_{min}$ these regions reduce to four single points
on the rays $\theta=0,\pi/2,\pi, 3\pi/2$, at the same radius 
$k=k_R[2/(1-{\cal D}^2+\sqrt{1-{\cal D}^2})]^{1/2}$. 
This is in contrast to the pure Rashba
case (${\cal D}=0$), where the bottom of the lower band is a circle of radius $k=k_R$ at 
$\varepsilon_F=-\varepsilon_R/2$.

\begin{figure}
    \centering
    \includegraphics[scale=0.6]{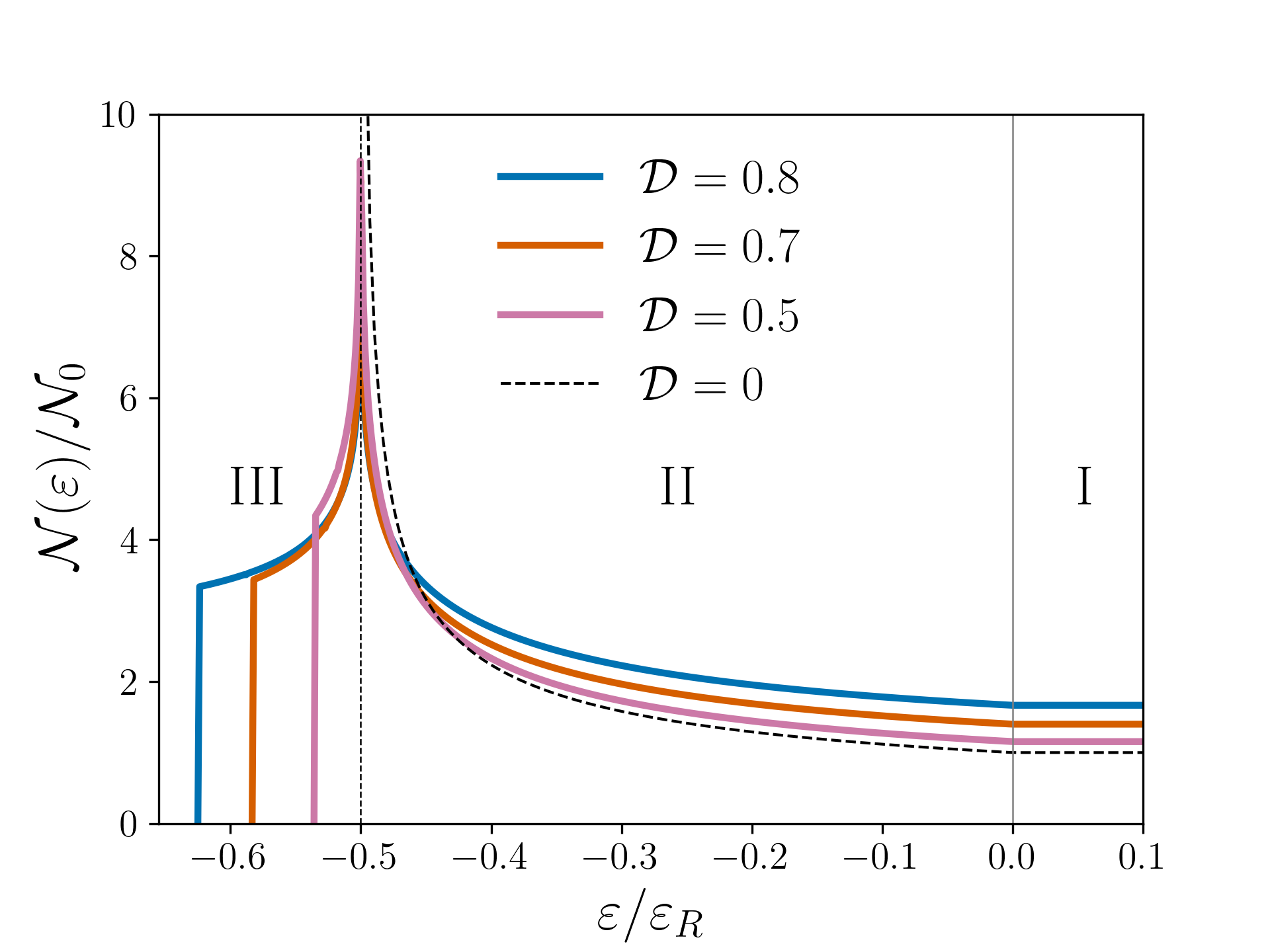}  
    \caption{Normalized DOS of a 2D $d$-wave altermagnet with RSOI for several values of the strength parameter $\mathcal{D}$. Three intervals (I, II, and III)
    are distinguished for the Fermi energy level position (see text).
    The dashed curve corresponds to the Rashba gas case $\mathcal{D}=0$.}
    \label{DOS}
\end{figure}

The density of states (DOS) follows this topological change accordingly, reading as: 
${\cal N}(\varepsilon)={\cal N}_I(\varepsilon)\Theta(\varepsilon)+
{\cal N}^<(\varepsilon)\Theta(-\varepsilon)$, where
${\cal N}^<(\varepsilon)={\cal N}_{II}(\varepsilon)\Theta(\varepsilon+\varepsilon_R/2)+
{\cal N}_{III}(\varepsilon)\Theta(-\varepsilon-\varepsilon_R/2)
\Theta(\varepsilon-\varepsilon_{min})$;
$\Theta(x)$ is the Heaviside unit step function. For positive energy $\varepsilon$,
${\cal N}_I(\varepsilon)={\cal N}_0/\sqrt{1-{\cal D}^2}$, where
${\cal N}_0=m/\pi\hbar^2$ is the DOS of the spin degenerate 2DEG (${\cal D}=0, \alpha=0$),
and for negative energy
\begin{equation} \label{DOS<}
{\cal N}^<(\varepsilon)=\frac{{\cal N}_0}{2\pi}\int_{\theta_1(\varepsilon)}^{\theta_2(\varepsilon)}\!\frac{d\theta}
{1-{\cal D}^2\cos^22\theta}\frac{\varepsilon_R+\varepsilon {\cal D}^2\cos^22\theta}
{\sqrt{\varepsilon_R^2+2\varepsilon\varepsilon_R+(\varepsilon{\cal D}\cos2\theta)^2}}\,.
\end{equation}
The angular region of integration $[\theta_1,\theta_2]$ is $[0,2\pi]$ for
$\varepsilon$ in the interval II,
and it is the sector $[-\theta^*,\theta^*]$ for $\varepsilon$ within the interval III
where $2\theta^*=\arccos(\sqrt{2\varepsilon_R(|\varepsilon|-\varepsilon_R/2)/{\cal D}^2\varepsilon^2})$ (see Fig.\,\ref{disjoint}); in the latter case, the integral 
(\ref{DOS<}) should be multiplied by four to include the contributions of all 
non-overlapping sectors containing the Fermi surface.
We obtain
\begin{equation} \label{NII}
    \mathcal{N}_{II}(\varepsilon)=\frac{\mathcal{N}_0}{2\pi}\frac{4(\varepsilon+\varepsilon_R)}{\sqrt{\varepsilon_R^2+2\varepsilon\varepsilon_R+(\varepsilon\mathcal{D})^2}}\left[\frac{1}{1-\mathcal{D}^2}\ \Pi\left(-\frac{\mathcal{D}^2}{1-\mathcal{D}^2},\,\zeta(\varepsilon)\right)-\frac{\varepsilon}{\varepsilon+\varepsilon_R}  K\left(\zeta(\varepsilon)\right)\right]\,,
\end{equation}
and 
\begin{eqnarray} \label{NIII}
    \mathcal{N}_{III}(\varepsilon)&=&4\,\frac{\mathcal{N}_0}{2\pi}\frac{\varepsilon+\varepsilon_R}{2\sqrt{\varepsilon_R^2+2\varepsilon\varepsilon_R+(\varepsilon\mathcal{D})^2}}\Bigg\{\frac{1}{1-\mathcal{D}^2}\Bigg[\Pi\left(-\frac{\mathcal{D}^2}{1-\mathcal{D}^2};\,2\theta^\ast \Bigg|\,\zeta(\varepsilon)\right)-\Pi\left(-\frac{\mathcal{D}^2}{1-\mathcal{D}^2};\,-2\theta^\ast \Bigg|\,\zeta(\varepsilon)\right)\Bigg]  \\
    && \hspace*{12cm} -\frac{2\varepsilon}{\varepsilon+\varepsilon_R}  F\left(2\theta^\ast,\,\zeta(\varepsilon)\right)\Bigg\}, \nonumber
\end{eqnarray}
where $K(m)$ and $\Pi(n,\,m)$ are the complete elliptic integrals of the first kind and of the third kind, respectively, $\zeta(\varepsilon)=(\varepsilon\mathcal{D})^2/(\varepsilon_R^2+2\varepsilon\varepsilon_R+
(\varepsilon\mathcal{D})^2)$,
and $F(\varphi,\,m)$ and $\Pi(n;\,\varphi|m)$ are the incomplete elliptic integrals of the first kind and of the third kind \cite{Abramowitz}. 
Fig.\,\ref{DOS} displays the DOS function ${\cal N}(\varepsilon)$
for the 2D $d$-wave altermagnet and for the pure Rashba case. As is known \cite{Winkler}, 
the DOS of the Rashba gas is given by
${\cal N}(\varepsilon)={\cal N}_0\left(\Theta(\varepsilon)+\Theta(-\varepsilon)\Theta(\varepsilon+\varepsilon_R/2)
\sqrt{\frac{\varepsilon_R/2}{\varepsilon+\varepsilon_R/2}}\,\right)$, which 
presents a high-order van Hove singularity \cite{Yuan2019} with a
power-law divergence near the bottom of the band $\varepsilon\gtrsim -\varepsilon_R/2$.
In contrast, the DOS of the Rashba altermagnet takes a finite value at
the bottom of the band $\varepsilon=\varepsilon_{min}$, but
diverges logarithmically at $\varepsilon=-\varepsilon_R/2$,
\begin{equation}
{\cal N}^<(\varepsilon)=\frac{2{\cal N}_0}{\pi{\cal D}}\times\left\{\begin{array}{ll}
\ln\left({\cal D}^2\,\frac{\varepsilon_R/2}{\varepsilon+\varepsilon_R/2}\right)
+ f_{II}(\zeta(-\varepsilon_R/2^+)), &
\varepsilon\gtrsim -\varepsilon_R/2 \\
\ln\left(4{\cal D}^2\,\frac{\varepsilon_R/2}{|\varepsilon+\varepsilon_R/2|}\right)
+ f_{III}(\zeta(-\varepsilon_R/2^-)), &
\varepsilon\lesssim -\varepsilon_R/2\ , 
\end{array}\right.
\end{equation}
where the functions $f_{II}$ and $f_{III}$ collect non divergent terms in the
vicinity of $\varepsilon=-\varepsilon_R/2$.



\section{Dynamical spin Edelstein response}\label{Sec3}

\subsection{Spin Edelstein susceptibility tensor}

The spin Edelstein effect (SEE) refers to the spin magnetization induced by an
externally applied electric field. Within a linear response theory framework,
it is characterized by the spin-velocity susceptibility using the Kubo formula
\cite{TheorySEE,SpinPolEng}
\begin{equation} \label{Kubo_gamma}
\gamma_{ij}(\omega)=ie\hbar\sum_{\bf k}\sum_{\lambda,\lambda'}\frac{f_{\lambda\lambda'}}{\varepsilon_{\lambda,\lambda'}}\,
\frac{\sigma^i_{\lambda\lambda'}v^j_{\lambda'\lambda}}{\hbar\tilde{\omega}+\varepsilon_{\lambda\lambda'}} \,,
\end{equation}
where we have allowed a frequency-dependent spin polarization density
$S_i(t)=S_i^{\omega}e^{-i\omega t}=(\hbar/2)\gamma_{ij}(\omega)E_j^{\omega}e^{-i\omega t}$ (sum over repeated indices is implied)
caused by a linearly polarized, spatially homogeneous, periodic time-varying electric field of frequency $\omega$, 
${\bf E}(\omega)=(E^{\omega}_x{\bf \hat{x}}+E^{\omega}_y{\bf\hat{y}})e^{-i\omega t}e^{\eta t}\, 
(\eta\to 0^+)$. In \eqref{Kubo_gamma},
the first Cartesian index refers to the direction of the induced spin polarization,
$i=x,y,z$, while the second refers to the direction of the in-plane electric field, $j=x,y$. 
We have also defined
 $\tilde{\omega}=\omega+i\eta$, and the sum in ${\bf k}$ is written as an integral in the continuum limit, $\sum_{\bf k}(\cdots)=\frac{1}{(2\pi)^2}\int d^2k(\cdots)$. Here $\lambda$ is a band index, and the energy difference 
$\varepsilon_{\lambda\lambda'}=\varepsilon_{\lambda}-\varepsilon_{\lambda'}$.
$\sigma^i_{\lambda\lambda'}=\langle\lambda|\hat{\sigma}_i|\lambda'\rangle$ are spin matrix elements,  $v^j_{\lambda'\lambda}=\langle\lambda'|\hat{v}_j|\lambda\rangle$ are velocity matrix elements, where
 $|\lambda\rangle$ is the eigenspinor of a state with wave vector ${\bf k}$ 
 in the band $\lambda$, $\hat{v}_j$
 is a component of the velocity operator ${\bf\hat{v}}=\nabla_{\bf k}H({\bf k})/\hbar$, and
 $-e$ is the electron charge. Here we shall consider the intrinsic effect only and treat
$f_{\lambda}=f(\varepsilon_{\lambda})$ as the equilibrium Fermi–Dirac distribution function for the band $\varepsilon_{\lambda}$, where $f_{\lambda\lambda'}=f_{\lambda}-f_{\lambda'}$.
 
The spin Edelstein susceptibility (\ref{Kubo_gamma}) can be separated into intraband ($\lambda=\lambda'$) and interband ($\lambda\neq\lambda'$)
contributions $\gamma_{ij}(\omega)=\gamma^{intra}_{ij}(\omega)+\gamma^{inter}_{ij}(\omega)$, 
where
\begin{eqnarray} 
\gamma_{ij}^{intra}(\omega)&=&\frac{ie}{\tilde{\omega}}\sum_{\lambda}\sum_{\bf k}\left(\frac{\partial f(\varepsilon_{\lambda})}{\partial\varepsilon_{\lambda}}\right)
\langle\lambda|\hat{\sigma}_i|\lambda\rangle\langle\lambda|\hat{v}_j|\lambda\rangle, \label{intra1} \\
\gamma_{ij}^{inter}(\omega)&=&e\hbar\sum_{\bf k}\sum_{\lambda,\lambda'}f_{\lambda\lambda'}\,\frac{\bar{T}^{ji}_{\lambda'\lambda}}{\hbar\tilde{\omega}+\varepsilon_{\lambda\lambda'}}\,, \label{inter1}
\ \ \ \ (\lambda\neq\lambda') \,.   
\end{eqnarray}
The interband part is expressed in terms of the Zeeman quantum geometric tensor
(Zeeman QGT) given by \cite{LXiangPRLb,LXiangPRLa}
\begin{equation}
\bar{T}^{ij}_{\lambda\lambda'}=-i\frac{v^i_{\lambda\lambda'}\sigma^j_{\lambda'\lambda}}{\varepsilon_{\lambda\lambda'}} \,, \ \ \ (\lambda\neq\lambda')\,.
\end{equation}
This spin$-$velocity tensor is a complex gauge-invariant quantity, 
which, by analogy to the conventional velocity$-$velocity QGT \cite{RMP_Nagaosa}
is written as
\begin{equation}
\bar{T}^{ij}_{\lambda\lambda'}({\bf k})=\bar{g}^{ij}_{\lambda\lambda'}({\bf k})-\frac{i}{2}\bar{\Omega}^{ij}_{\lambda\lambda'}({\bf k}),
\end{equation}
where $\bar{\Omega}^{ij}_{\lambda\lambda'}\equiv -2\text{Im}(\bar{T}^{ij}_{\lambda\lambda'})$ is the (local) {\it Zeeman Berry curvature} (ZBC) and
$\bar{g}^{ij}_{\lambda\lambda'}\equiv \text{Re}(\bar{T}^{ij}_{\lambda\lambda'})$ is the (local) {\it Zeeman quantum metric} (ZQM). 
In the case of a generic 2D two-band model $H({\bf k})=\epsilon_0({\bf k})\sigma_0+\boldsymbol{\sigma}\cdot {\bf d}({\bf k})$,
\begin{eqnarray}
\bar{g}^{ij}_{\lambda,-\lambda}({\bf k})&=&\frac{1}{\hbar}\frac{1}{2d^2}\left({\bf d}\times\frac{\partial{\bf d}}{\partial k_i}\right)_j\ , \label{ZQM} \\
\bar{\Omega}^{ij}_{\lambda,-\lambda}({\bf k})&=&\frac{\lambda}{\hbar}\frac{1}{d^3}\left[\left({\bf d}\times\frac{\partial{\bf d}}{\partial k_i}\right)\times{\bf d}\right]_j\,, \label{ZBC}
\end{eqnarray}
where $d({\bf k})=|{\bf d}({\bf k})|$.

At zero temperature, from (\ref{intra1}) and (\ref{inter1}),
\begin{eqnarray}
\gamma_{ij}^{intra}(\omega)&=&-\frac{ie}{\hbar\tilde{\omega}}\sum_{\lambda=\pm}\lambda\int\!\!\frac{d^2k}{(2\pi)^2}
\frac{d_i}{d}\left(\frac{\partial\varepsilon_{\lambda}}{\partial k_j}\right)\delta(\varepsilon_{\lambda}({\bf k})-\varepsilon_F)\,, \label{IntraB2} \\
\gamma_{ij}^{inter}(\omega)&=&e\hbar\int'\!\!\frac{d^2k}{(2\pi)^2}\,\frac{4d({\bf k})\,\bar{g}^{ji}({\bf k})-i\hbar\tilde{\omega}\,\bar{\Omega}^{ji}({\bf k})}{(\hbar\tilde{\omega})^2-(2d({\bf k}))^2}\,, \label{InterB2}
\end{eqnarray}
where $\bar{\Omega}^{ji}=\bar{\Omega}^{ji}_{+-}$ and $\bar{g}^{ji}=\bar{g}^{ji}_{+-}$ (see (\ref{ZQM}) and(\ref{ZBC})). 
The prime on the integral means integration over the ${\bf k}$-region satisfying $\varepsilon_-({\bf k})\leqslant\varepsilon_F\leqslant\varepsilon_+({\bf k})$. 
In interval I, this is the region between $k_F^+(\theta)$ and $k_F^-(\theta)$
with $\theta\in[0,2\pi]$ (Fig.~\ref{FermiCont}a); in interval II, it is the region between $k_{F,-}^-(\theta)$ and 
$k_{F,+}^-(\theta)$ with $\theta\in[0,2\pi]$ (Fig.~\ref{FermiCont}b,c), and in interval III, it is the region
between $k_{F,-}^-(\theta)$ and $k_{F,+}^-(\theta)$ restricted to the angular sectors
defined by (\ref{abscos}) (Fig.~\ref{disjoint} and Fig.~\ref{FermiCont}d).


\begin{table}[h]
\centering
\renewcommand{\arraystretch}{1.3}
\begin{tabular}{c c c}
\toprule
\ \ \ & $\bar{g}^{ij}_{\lambda\lambda'}({\bf k})$ &   $\bar{\Omega}^{ij}_{\lambda\lambda'}({\bf k})$ \\
\midrule
$\mathcal{P}$ & odd & odd \\
$\mathcal{T}$ & odd & even \\
$\mathcal{PT}$ & even & odd \\
$C_{2z}$ & even ($j\neq z$) & even ($j\neq z$)\\
$C_{2z}$ & odd ($j=z$) & odd ($j= z$)\\

\bottomrule
\end{tabular}
\caption{Symmetry properties of the ZQM and ZBC in momentum space, due to inversion ($\mathcal{P}$), time-reversal ($\mathcal{T}$), $\mathcal{P}\mathcal{T}$, and two-fold rotation ($C_{2z}$) symmetries.}
\label{Table01}
\end{table}

\subsection{Symmetries of the model}\label{SymmModel}

In order to investigate the structure of the interband Edelstein susceptibility tensor for our model Hamiltonian \eqref{Ham},
we discuss the action of some relevant symmetry operations on the momentum-dependent Zeeman QGT.

We first note that $\bar{g}^{ij}_{\lambda\lambda'}({\bf k})=\bar{g}^{ij}_{\lambda'\lambda}({\bf k})$ and $\bar{\Omega}^{ij}_{\lambda\lambda'}({\bf k})=-\bar{\Omega}^{ij}_{\lambda'\lambda}({\bf k})$.
Under $\mathcal{T}$ operation, $\hat{\mathcal{T}}\sigma_{\lambda\lambda'}^i=-\sigma_{\lambda'\lambda}^i$ and $\hat{\mathcal{T}}r_{\lambda\lambda'}^i=r_{\lambda'\lambda}^i$, where $r_{\lambda\lambda'}^i=-iv_{\lambda\lambda'}^i/\varepsilon_{\lambda\lambda'}$, and under $\mathcal{P}$ operation,
$\hat{\mathcal{P}}\sigma_{\lambda\lambda'}^i=\sigma_{\lambda\lambda'}^i$ and $\hat{\mathcal{P}}r_{\lambda\lambda'}^i=-r_{\lambda\lambda'}^i$, where $\hat{\mathcal{P}}$ is the spatial inversion operator.
Therefore, in a $\mathcal{T}(\mathcal{P})$-invariant system, $\bar{\Omega}^{ij}_{\lambda\lambda'}({\bf k})$ is an even (odd) function in momentum, whereas it vanishes in systems that are invariant under both $\mathcal{T}$ and $\mathcal{P}$.
In contrast, the ZQM is odd in momentum under either $\mathcal{P}$ or $\mathcal{T}$ symmetry individually, and hence even in a $\mathcal{PT}$-invariant system.
Another relevant operation is a two-fold rotation around the $z$-axis $C_{2z}=-(C_{4z}\hat{\mathcal{T}})^2$.
In a $C_{2z}$-invariant system, such as the model Hamiltonian (\ref{Ham}), $\bar{g}^{ij}_{\lambda\lambda'}({\bf k})$ and $\bar{\Omega}^{ij}_{\lambda\lambda'}({\bf k})$ are  even (odd) functions in momentum for $j\neq z$ ($j=z$).
All of these symmetry properties are summarized in Table \ref{Table01}.
We note that in the absence of RSOI but finite $\mathcal{D}$ in \eqref{Ham}, $\bar{g}^{ij}_{\lambda\lambda'}({\bf k})$ and $\bar{\Omega}^{ij}_{\lambda\lambda'}({\bf k})$  ($j\neq z$) vanish due to $\mathcal{P}$ and $C_{2z}$ symmetries. 
Likewise, the $U(1)$ spin-rotation symmetry makes $\sigma^z_{\lambda\lambda'}=0$, leading to $\bar{g}^{iz}_{\lambda\lambda'}({\bf k})=0$ and $\bar{\Omega}^{iz}_{\lambda\lambda'}({\bf k})=0$.  
Thus, in the absence of RSOI, the Zeeman QGT vanishes.
On the other hand, taking $\mathcal{D}=0$ but finite RSOI leads to a vanishing $\bar{g}^{ij}_{\lambda\lambda'}({\bf k})$ ($j\neq z$) and $\bar{\Omega}^{iz}_{\lambda\lambda'}({\bf k})$ due to $\mathcal{T}$ and $C_{2z}$ symmetries, while the other components can be nonzero.

As $C_{4z}\hat{\mathcal{T}}$ is a symmetry operation of the altermagnet Hamiltonian \eqref{Ham}, the Zeeman QGT obeys $(C_{4z}\hat{\mathcal{T}})\bar{T}^{ij}_{\lambda\lambda'}({\bf k})=\bar{T}^{ij}_{\lambda\lambda'}(C_{4z}\hat{\mathcal{T}}{\bf k})=\bar{T}^{ij}_{\lambda\lambda'}(k_y\,,-k_x)$, which leads to symmetry properties of the ZQM and ZBC in momentum space as $-\sum_{a,b=x,y}
\varepsilon_{iaz}\varepsilon_{jbz}\,
\bar g^{ab}_{\lambda\lambda'}(k_x,k_y)+\delta_{jz}\sum_{a=x,y}
\varepsilon_{iaz}\,
\bar g^{az}_{\lambda\lambda'}(k_x,k_y)=\bar g^{ij}_{\lambda\lambda'}(k_y,-k_x)$, and $\sum_{a,b=x,y}
\varepsilon_{iaz}\varepsilon_{jbz}\,
\bar \Omega^{ab}_{\lambda\lambda'}(k_x,k_y)-\delta_{jz}\sum_{a=x,y}
\varepsilon_{iaz}\,
\bar \Omega^{az}_{\lambda\lambda'}(k_x,k_y)=\bar \Omega^{ij}_{\lambda\lambda'}(k_y,-k_x)$, respectively.
We also note that the composite symmetry operations $\mathcal{M}_{xz}\hat{\mathcal{T}}$ and $\mathcal{M}_{yz}\hat{\mathcal{T}}$ of \eqref{Ham}, where $\mathcal{M}_{xz}$ ($\mathcal{M}_{yz}$) is a mirror reflection through the $xz\,(yz)$-plane, yield additional symmetry properties as 
$M_{\pm}^{ij}\bar{g}^{ij}(k_x,\,k_y)=\bar{g}^{ij}(\mp k_x,\,\pm k_y)$ and $-M_{\pm}^{ij}\bar{\Omega}^{ij}(k_x,\,k_y)=\bar{\Omega}^{ij}(\mp k_x,\,\pm k_y)$, where the upper (lower) sign refers to the effect of the symmetry operation $\mathcal{M}_{xz}\mathcal{T}$ ($\mathcal{M}_{yz}\mathcal{T}$), and we have defined $M_{\pm}^{ij}=(\delta_{ix}-\delta_{iy})(\delta_{jx}-\delta_{jy})\pm(\varepsilon_{ixj}+\varepsilon_{iyj})$.
Such properties determine the structure of the tensor $\gamma^{inter}_{ij}(\omega)$ for the model Hamiltonian $\eqref{Ham}$. 
We find that the longitudinal components 
are determined exclusively by the ZQM and $\gamma^{inter}_{yy}(\omega)=-\gamma^{inter}_{xx}(\omega)$, while the off-diagonal components are determined exclusively by the ZBC and $\gamma^{inter}_{yx}(\omega)=-\gamma^{inter}_{xy}(\omega)$. 
Here $\gamma^{inter}_{zj}(\omega)$ always  vanishes.


\begin{table}[h]
\centering
\renewcommand{\arraystretch}{1.3}
\begin{tabular}{l|l}
\toprule
\ \ \ \ ZQM & \ \ \ \ \ \ \ \ ZBC \\
\midrule
$\bar{g}^{xx}({\bf k})=-(\alpha{\cal D}/2\hbar d^2)(\hbar^2k^2/2m)$ & $\hbar d^3\bar{\Omega}^{xx}({\bf k})=-\alpha^3k_xk_y-\alpha{\cal D}h({\bf k})(\hbar^2k_xk_y/m)$ \\
$\bar{g}^{yy}({\bf k})=-\bar{g}^{xx}({\bf k})$ & $\hbar d^3\bar{\Omega}^{yy}({\bf k})=\alpha^3k_xk_y-\alpha{\cal D}h({\bf k})(\hbar^2k_xk_y/m)$\\
$\bar{g}^{xy}({\bf k})=-(\alpha{\cal D}/2\hbar d^2)(\hbar^2k_xk_y/m)$ & $\hbar d^3\bar{\Omega}^{xy}({\bf k})=-\alpha^3k_y^2+\alpha{\cal D}h({\bf k})(\hbar^2k^2/2m)$ \\
$\bar{g}^{yx}({\bf k})=-\bar{g}^{xy}({\bf k})$ & $\hbar d^3\bar{\Omega}^{yx}({\bf k})=\alpha^3k_x^2+\alpha{\cal D}h({\bf k})(\hbar^2k^2/2m)$\\
$\bar{g}^{xz}({\bf k})=-\alpha^2k_y/2\hbar d^2$ & $\hbar d^3\bar{\Omega}^{xz}({\bf k})=\alpha^2{\cal D}k_x[\hbar^2(k_x^2+3k_y^2)/2m]$ \\
$\bar{g}^{yz}({\bf k})=\alpha^2k_x/2\hbar d^2$ & $\hbar d^3\bar{\Omega}^{yz}({\bf k})=-\alpha^2{\cal D}k_y[\hbar^2(3k_x^2+k_y^2)/2m]$ \\

\bottomrule
\end{tabular}
\caption{Real ($\bar{g}$) and imaginary ($\bar{\Omega}$) parts of the Zeeman QGT of the 2D Rashba altermagnet model, which determine the interband contribution
to the spin Edelstein susceptibility. 
In the absence of RSOI ($\alpha=0$) the tensor vanishes, leading to a null interband susceptibility. In the second column $h({\bf k})={\cal D}\hbar^2(k_x^2-k_y^2)/2m$ is the altermagnetic energy contribution.}
\label{Table4}
\end{table}

\subsection{Intraband response}

\begin{figure}
    \centering
    \includegraphics[scale=0.46]{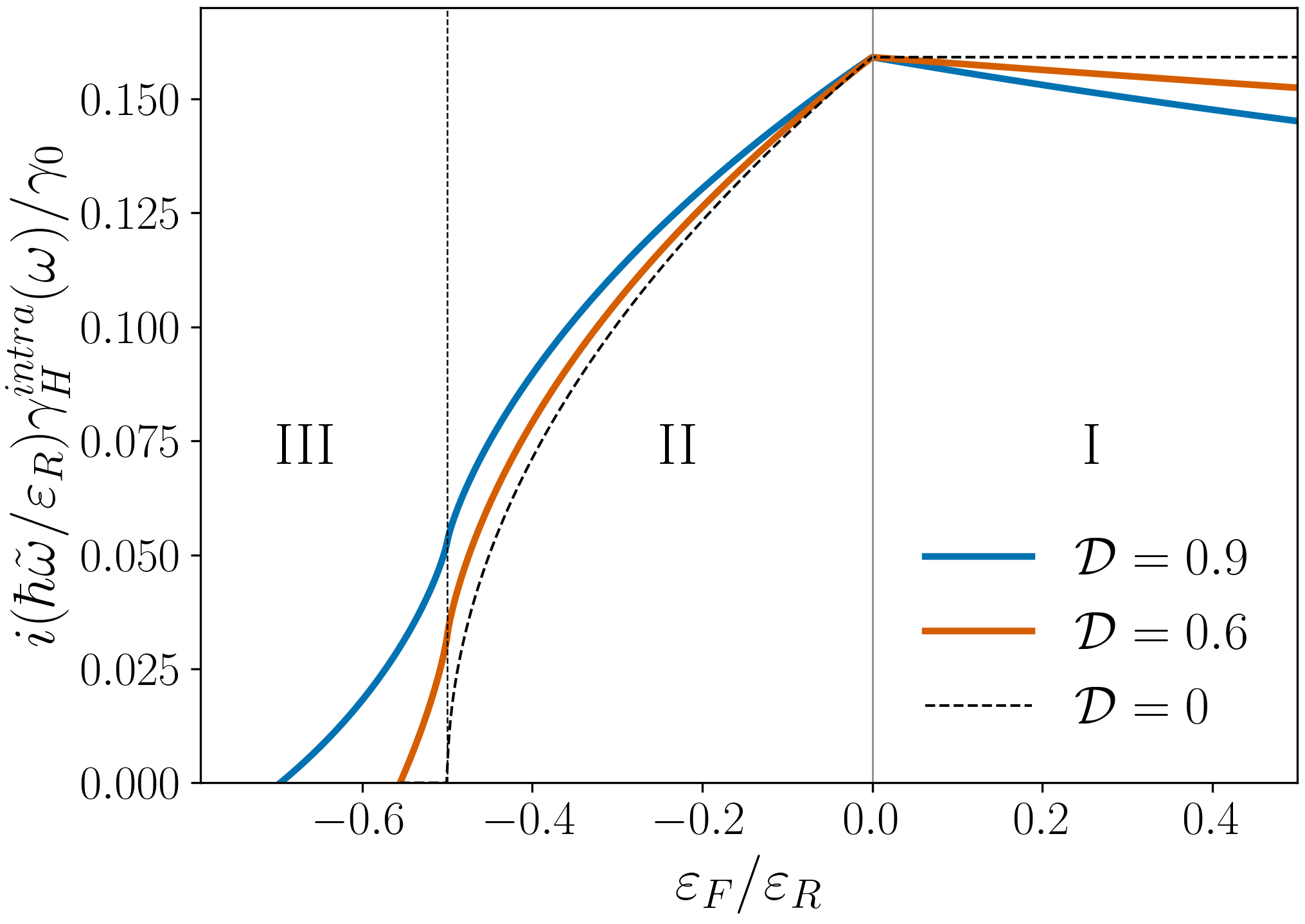}
    \caption{Fermi energy dependence of the intraband transverse spin Edelstein susceptibility $\gamma_{ij}^{intra}(\omega)=\epsilon_{zij}\gamma_H^{intra}(\omega)$
    for several values of the altermagnet parameter ${\cal D}$. 
   }
    \label{IntraGamma}
\end{figure}

For our model, it is verified from (\ref{IntraB2}) that 
$\gamma^{intra}_{zj}(\omega)=0$, given the symmetry of the integrand and the
region of integration for each interval; the factor $d_z/d=h({\bf k})/d$ is
even in $k$, while $\partial\varepsilon_{\lambda}/\partial k_j$ is an odd function.
The in-plane part of the intraband tensor becomes off-diagonal.
$\gamma^{intra}_{ij}(\omega)=\epsilon_{zij}\gamma_H^{intra}(\omega)$
($\epsilon_{ijk}$ is the antisymmetric Levi-Civita symbol).
Figure \ref{IntraGamma} displays this Hall response as a function
of the Fermi energy for several values of the altermagnetic parameter ${\cal D}$. 
For ${\cal D}=0$, we recover the known result for Rashba systems \cite{TheorySEE} .
In interval I, it is
a decreasing function of $\varepsilon_F$ and ${\cal D}$.
At $\varepsilon_F=0$, its magnitude becomes independent of ${\cal D}$, $\gamma_H^{intra}(\omega)=i\gamma_0(\varepsilon_R/2\pi\hbar\omega)$.
In interval II, the behavior looks very similar to that of the pure Rashba case.
In interval III, a change in curvature is observed, with a magnitude that increases as ${\cal D}$ and vanishes for $\mathcal{D}=0$. We shall use the quantity $\gamma_0=e/\alpha$ as a normalization factor.

\subsection{Interband response}

Regarding the interband contribution, 
using the Plemelj formula $1/(x+i0^+)=\mathcal{P}(1/x)-i\pi\delta(x)$ in
(\ref{InterB2}), we rewrite it as
\begin{eqnarray}
\text{Re}\gamma_{ij}^{inter}(\omega)&=& e\hbar \mathcal{P}\int'\frac{d^2k}{(2\pi)^2}
\frac{4d(\mathbf{k})\,\bar{g}^{ji}(\mathbf{k})}
{(\hbar\omega)^2-(2d(\mathbf{k}))^2}
-\frac{\pi}{2}e\hbar
\int'\frac{d^2k}{(2\pi)^2}
\bar{\Omega}^{ji}(\mathbf{k})
\,\delta\!\left(\hbar\omega-2d(\mathbf{k})\right)\,, \label{ReGamma}
\\
\text{Im}\gamma_{ij}^{inter}(\omega)
&=&
-e\hbar^2\omega\mathcal{P}
\int'\frac{d^2k}{(2\pi)^2}
\frac{\bar{\Omega}^{ji}(\mathbf{k})}
{(\hbar\omega)^2-(2d(\mathbf{k}))^2}
-\pi e\hbar
\int'\frac{d^2k}{(2\pi)^2}
\bar{g}^{ji}(\mathbf{k})
\,\delta\left(\hbar\omega-2d(\mathbf{k})\right)\,, \label{ImGamma}
\end{eqnarray}
where $\mathcal{P}$ denotes the principal value integral.

\begin{figure}
    \centering
    \includegraphics[scale=0.5]{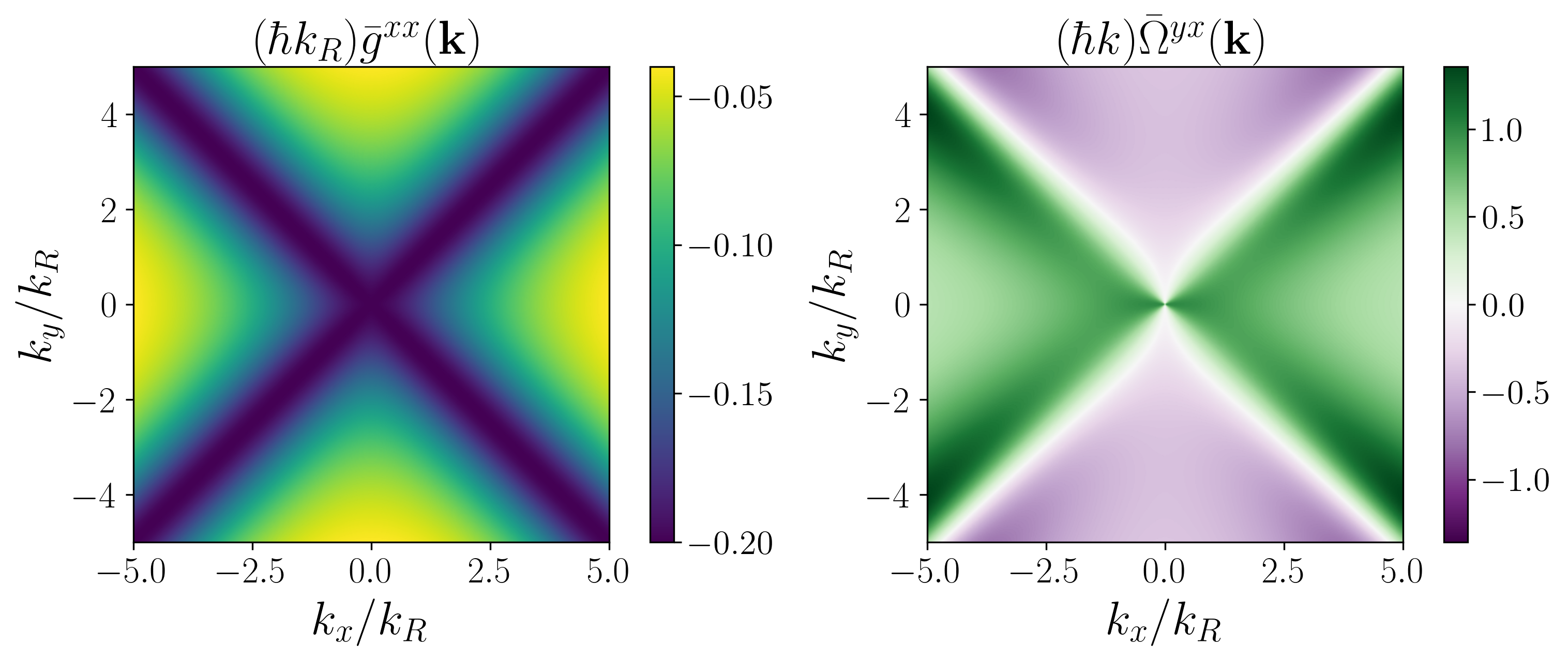}
    \caption{Left: Longitudinal ZQM component $\bar{g}^{xx}({\bf k})$.
    Right: Transverse ZBC component $\bar{\Omega}^{yx}({\bf k})$, multiplied by $\hbar k$
    to remove its divergence at $k=0$. These quantities determine the interband contribution to the longitudinal and transverse spin Edelstein response, respectively.
    We use $\mathcal{D}=0.8$.}
    \label{ZeemanMetric}
\end{figure}

In the present case, ${\bf d}({\bf k})=\alpha(k_y{\bf\hat{x}}-k_x{\bf\hat{y}})+h({\bf k}){\bf\hat{z}}$. Note that in the absence of RSOI  ($\alpha=0$), ${\bf d}$ has
a fixed direction, ${\bf d}$ and $\partial{\bf d}/\partial k_i$ are collinear, and
${\bf d}\times\partial{\bf d}/\partial k_i=0$, so that the Zeeman QGT vanishes (see Eqs. (\ref{ZQM}) and (\ref{ZBC})), as expected. In turn, this implies the vanishing of the interband contribution to the spin Edelstein susceptibility (\ref{inter1}). In Table \ref{Table4} we show the analytical expressions of the ZQM and ZBC of the 2D Rashba $d$-wave altermagnet (\ref{Ham}).
Fig.\,\ref{ZeemanMetric} shows the longitudinal ZQM component $\bar{g}^{xx}$ and the transverse ZBC component $\bar{\Omega}^{yx}$ in momentum space. The symmetry properties discussed in Sec.\,\ref{SymmModel} are clearly seen. 
The maximum negative values of $\bar{g}^{xx}$ occur along the directions where the separation
between the bands $2d({\bf k})$ is minimum ($\cos^22\theta=0$, see Eq. (\ref{ladé})).
Thus, $\bar{g}^{xx}$ mainly probes the strength of interband transitions between the closest initial and final energy states. We note that the function $\bar{\Omega}^{yx}({\bf k})\approx
\cos^2\theta/\hbar k$ close to $k=0$. In Fig.~\ref{ZeemanMetric}, this divergence 
was removed by multiplying by $\hbar k$ without altering the symmetry of the function.
Away from the origin, the symmetry is mainly governed by the sign of $h({\bf k})\propto k_x^2-k_y^2$.



The calculations show that $\gamma_{zj}^{inter}(\omega)=0$,
$\gamma^{inter}_{xx}(\omega)=-\gamma_{yy}^{inter}(\omega)$, and 
$\gamma^{inter}_{xy}(\omega)=-\gamma^{inter}_{yx}(\omega)$ are valid in
the entire frequency range for any position of the Fermi level, as anticipated by the symmetry analysis in Sec. \ref{SymmModel}.

Before commenting on the frequency dependence of interband spin Edelstein 
susceptibility, we first consider its dc limit. At zero frequency we find
\begin{equation}
\gamma^{inter}_{xx}(0)=\gamma_0
\frac{{\cal D}}{2\pi}\left(\frac{\varepsilon_R}{4\varepsilon_R+2\varepsilon_F{\cal D}^2}\right)^{1/2}\,,
\end{equation}
in interval I,
\begin{equation}
    \gamma_{xx}^{inter}(0)=\gamma_0\frac{\mathcal{D}}{(2\pi)^2}\frac{4\varepsilon_F}{\sqrt{\varepsilon_R^2+2\varepsilon_F\varepsilon_R+(\varepsilon_F\mathcal{D})^2}} \left[ K\left(\zeta(\varepsilon_F)\right) + \frac{\varepsilon_R^2}{\mathcal{D}^2 \varepsilon_F^2 + 2\varepsilon_F\varepsilon_R} \Pi\left(\frac{\mathcal{D}^2 \varepsilon_F}{\mathcal{D}^2 \varepsilon_F + 2\varepsilon_R},\,\zeta(\varepsilon_F)\right) \right]\,,
\end{equation}
in interval II, and
\begin{eqnarray}
  \nonumber   \gamma_{xx}^{inter}(0)&=&\gamma_0\frac{\mathcal{D}}{(2\pi)^2}\frac{4\varepsilon_F}{
\sqrt{\varepsilon_R^2+2\varepsilon_F\varepsilon_R+(\varepsilon_F\mathcal{D})^2}}
\Bigg\{F\left(
2\theta^\ast,
\zeta(\varepsilon_F)
\right)+ \frac{\varepsilon_R^2}{2(\mathcal{D}^2\varepsilon_F^2+2\varepsilon_F\varepsilon_R)}\\
&&\hspace*{3cm}
\times\left[
\Pi\left(
\frac{\mathcal{D}^2\varepsilon_F}{\mathcal{D}^2\varepsilon_F+2\varepsilon_R};\,
2\theta^\ast\Bigg|
\zeta(\varepsilon_F)
\right)
-
\Pi\left(
\frac{\mathcal{D}^2\varepsilon_F}{\mathcal{D}^2\varepsilon_F+2\varepsilon_R};\,
-2\theta^\ast\Bigg|
\zeta(\varepsilon_F)
\right)
\right]
\Bigg\}\,,
\end{eqnarray}
in interval III. The function $\zeta(\varepsilon)$ is the same as that appearing in the
elliptic integrals (\ref{NII}) and (\ref{NIII}). 

\begin{figure}
    \centering
    \includegraphics[scale=0.496]{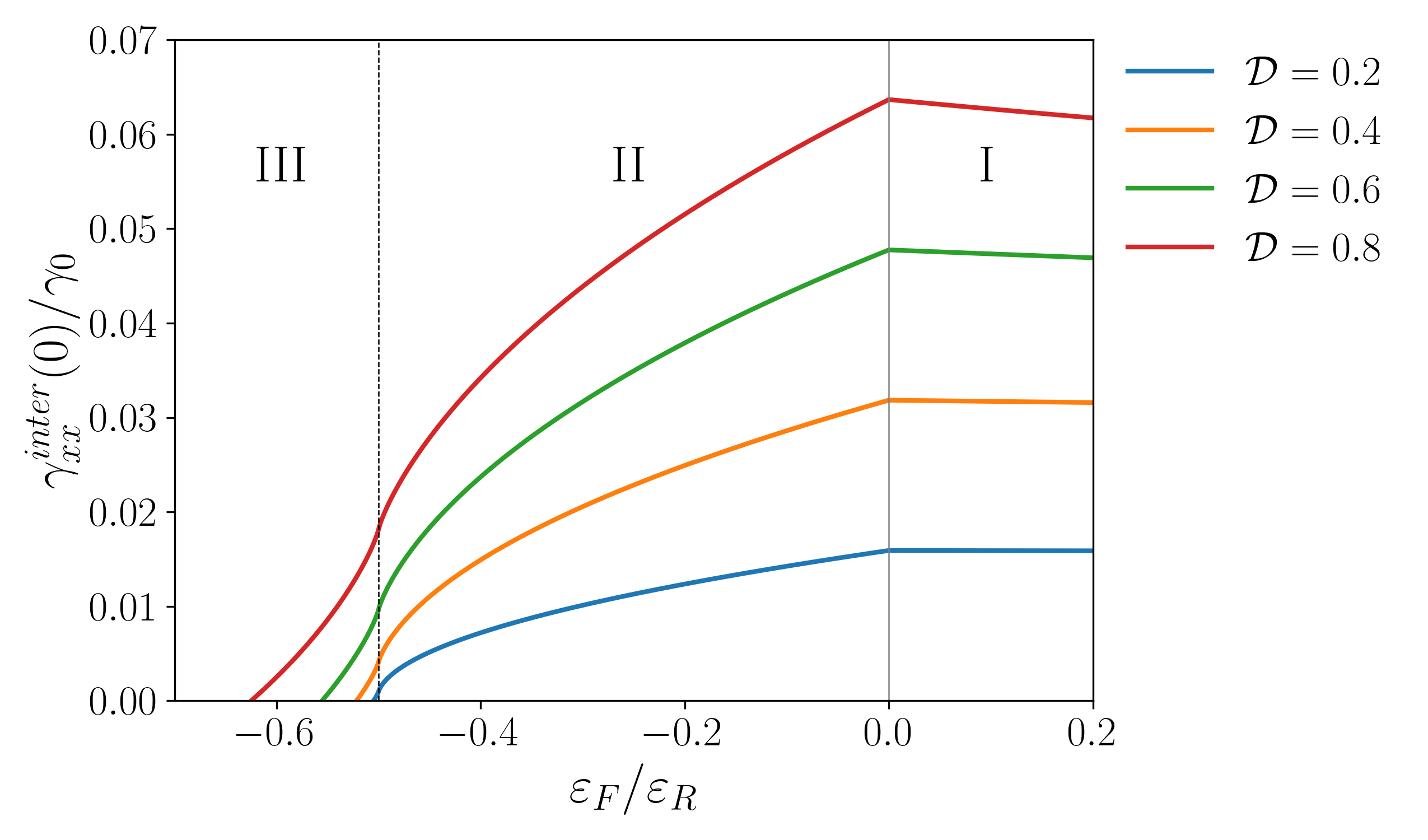}
    \caption{Interband longitudinal spin Edelstein susceptibility as a function of Fermi energy
    in the dc limit, for several values of the altermagnetic parameter
    $\mathcal{D}$.}
    \label{Inter_xx_Static}
\end{figure}

Fig.\,\ref{Inter_xx_Static} shows this longitudinal dc response for varying 
Fermi energy and several values of the altermagnetic strength parameter.
The maximum value appears at zero Fermi energy as a kink of magnitude ${\cal D}/4\pi$,
independent of the Rashba parameter $\alpha$. Away from this point, it decreases slowly
in region I but more rapidly for negative values of $\varepsilon_F$. 
We note that for the Rashba nonaltermagnetic phase ($\mathcal{D}=0$), this response vanishes regardless of the position of the Fermi energy.
From Eq. (\ref{ImGamma}), we obtain $\text{Im}\gamma^{inter}_{xy}(0)=0$ regardless
of the position of the Fermi level. For the corresponding real part, we find that it
also vanishes \cite{Nonlinear-anomalous,LXiangPRLb}, except at $\varepsilon_F=0$, where it takes the value $\text{Re}\gamma^{inter}_{xy}(0)=-\gamma_0/16$, independent of the parameter ${\cal D}$; 
this result arises from the integral involving the ZBC in (\ref{ReGamma}). 
Thus, $\gamma_{ij}^{inter}(0)$ ($i,j=x,y$) is diagonal when $\varepsilon_F\neq 0$, with a nonzero
off-diagonal component appearing only for $\varepsilon_F=0$.
Note that this finite Hall response at $\varepsilon_F=0$ is absent for gapped
systems under nonresonant driving frequency, such that $\hbar\omega< 2d({\bf k})$
for all ${\bf k}$, which is a dynamical regime without direct interband transitions \cite{LXiangPRLb,DESG}.

\begin{figure}
    \centering
    \includegraphics[scale=0.5]{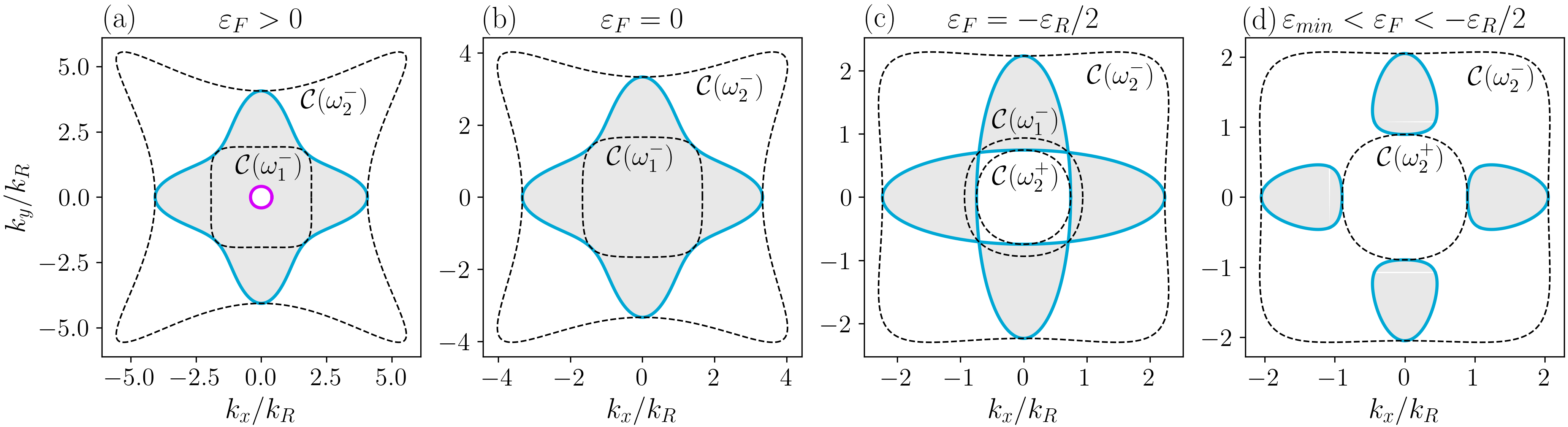}
    \caption{Fermi contours (solid lines) and resonance curves ${\cal C}(\omega)$ (black dashed) at some critical frequencies for several values of the Fermi energy. Note that the
    shown curves ${\cal C}(\omega)$ touch the Fermi lines tangentially.
    Shaded areas indicate the ${\bf k}$-region available for allowed vertical transitions.
    (a) Fermi contours in region I, defined by the equations $\varepsilon_+=\varepsilon_F$ (pink) and $\varepsilon_-=\varepsilon_F$ (blue). Only the resonance curves ${\cal C}(\omega_1^-)$ (inner) and ${\cal C}(\omega_2^-)$ (outer) touching the
    band $\varepsilon_-({\bf k})$ are shown.
    (b) At $\varepsilon_F=0$, only a single Fermi line (blue) develops, defined by the 
    $\varepsilon_-({\bf k})$ band. The depicted inner (outer) resonance curve is ${\cal C}(\omega_1^-)\,({\cal C}(\omega_2^-))$. (c) At $\varepsilon_F=-\varepsilon_R/2$, the Fermi contours (blue)
    defined by the band $\varepsilon_-({\bf k})$ touch each other along directions defined by
    $\cos2\theta=0$. Also shown are the
    set of resonance curves ${\cal C}(\omega_2^+)$ (inner), ${\cal C}(\omega_1^-=\omega_1^+)$, and ${\cal C}(\omega_2^-)$ (outer). (d) In region III, the Fermi contour breaks into four non-overlapping closed lines. The shown resonance curves are ${\cal C}(\omega_2^+)$ (inner) and 
    ${\cal C}(\omega_2^-)$ (outer).}
    \label{FermiCont}
\end{figure}

\begin{figure}
    \centering
    \includegraphics[scale=0.5]{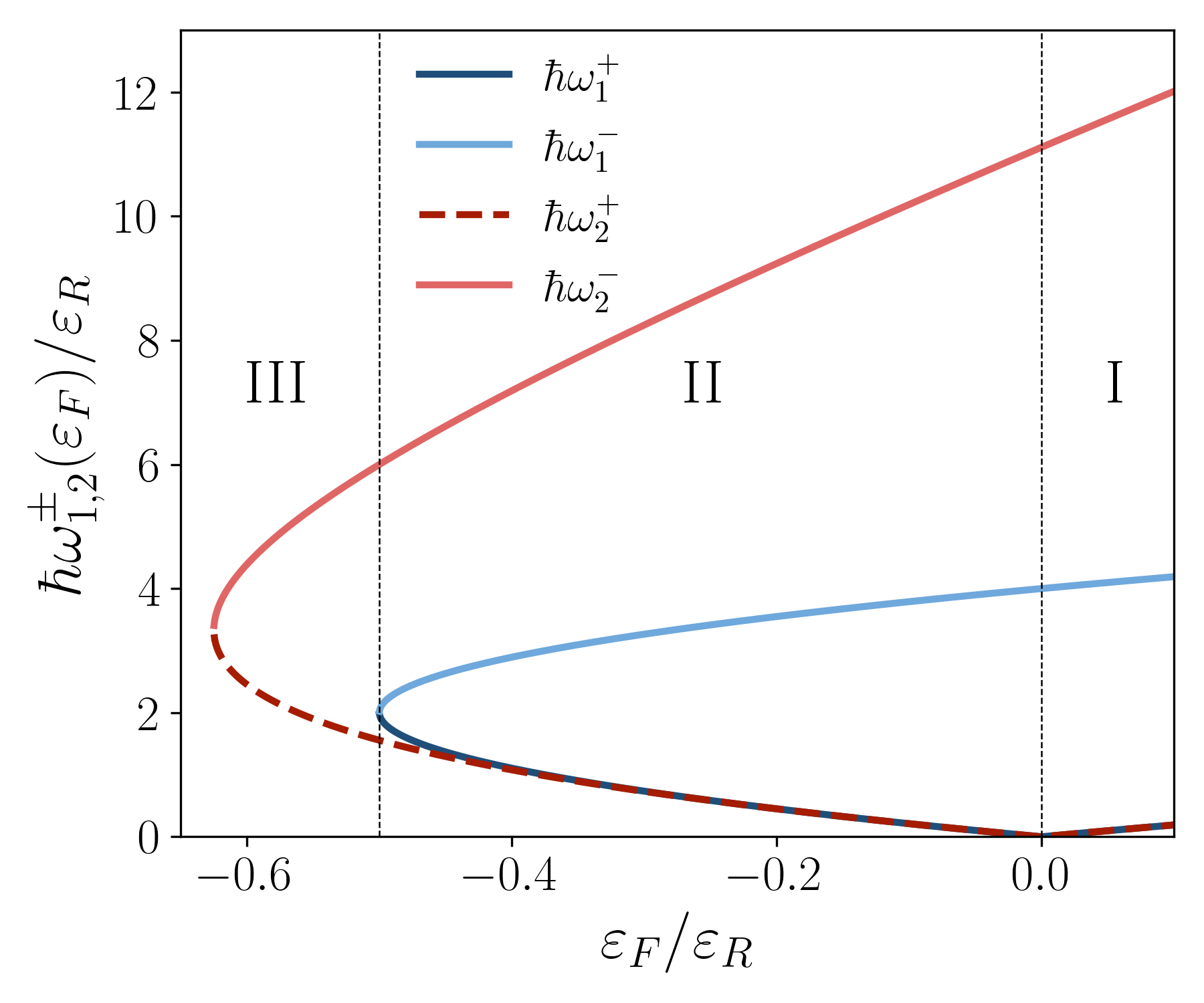}
    \caption{Critical frequencies which locate the van Hove singularities of the spin Edelstein
    susceptibility tensor of the Rashba altermagnet, as functions of the Fermi energy.
    Vertical dashed lines at $\varepsilon_F=0$ and $\varepsilon_F=-\varepsilon_R/2$ 
    indicate the borders between regions I and II, and between regions II and III, respectively.
    We take $\mathcal{D}=0.8$. For $\mathcal{D}=0$, $\hbar\omega_2^{\pm}=\hbar\omega_1^{\pm}$.}
    \label{CriticalF}
\end{figure}

To understand the spectral features of the response function $\gamma_{ij}^{inter}(\omega)$,
we first note that the last integrals in (\ref{ReGamma}) and (\ref{ImGamma}) are to be
performed under the restrictions $\varepsilon_-({\bf k})<\varepsilon_F<\varepsilon_+({\bf k})$
(Pauli blocking) and $2d({\bf k})=\hbar\omega$ (the `resonance curve'). 
The ${\bf k}$-region defined by the former condition is illustrated as shaded areas in Fig.~\ref{FermiCont}, bounded by the set of Fermi contours formed according to the location of 
the Fermi level (the same depicted in Fig.~\ref{Surf}). 
The second condition defines a curve ${\cal C}(\omega)=\{{\bf k}|2d({\bf k})=\hbar\omega\}$
whose shape depends on frequency.
Thus, for a given frequency, only those ${\bf k}$ vectors in a shaded area that lie on
the resonance curve will contribute to the mentioned integrals.
Extreme values of the energy difference function $2d({\bf k})$ arise for some ${\bf k}^c_F$ points of the resonance curve intersecting the Fermi contours. This poses an optimization problem with constraints: to find the extreme values of
the function $2d({\bf k})$ subject to the constraint $\varepsilon_{\lambda}({\bf k})=\varepsilon_F$. This, in turn, defines a set of critical frequencies 
$\hbar\omega_c=2d({\bf k}^c_F)$ at which the spectra develop characteristic features \cite{Topical}.
The resonance curves at critical frequencies touching the Fermi lines 
$\varepsilon_-({\bf k}^c_F)=\varepsilon_F$ are illustrated in Fig.~\ref{FermiCont}
as black dashed lines. The observed tangency between the resonance curve and the Fermi lines
in each case reflects the special character of the critical frequencies.
This resembles the spectral behavior of the joint density of states and optical conductivity
of two-band models \cite{Mojarro2021,Topical}. 
Four relevant energy difference functions can be identified:
$(i)\ 2d(k_F^+(\theta))=\varepsilon_F-\varepsilon_-(k_F^+(\theta))$ and $(ii)\ 
2d(k_F^-(\theta))=\varepsilon_+(k_F^-(\theta))-\varepsilon_F$ when $\varepsilon_F>0$,
and $(iii)\ 2d(k_{F,-}^-(\theta))=\varepsilon_+(k_{F,-}^-(\theta))-\varepsilon_F$
and $(iv)\ 2d(k_{F,+}^-(\theta))=\varepsilon_+(k_{F,+}^-(\theta))-\varepsilon_F$
when $\varepsilon_F<0$. These functions give the absorption edges
for a given $\theta$ and $\varepsilon_F$, according to the restrictions
imposed by Pauli blocking.
Identifying the symmetry directions $\theta_1=\pi/4,3\pi/4,5\pi/4$ or $7\pi/4$  and $\theta_2=0,\pi/2,\pi$ or $3\pi/2$, the resulting set of critical frequencies is:
%
in region I, $\hbar\omega^+_1=2d(k_F^+(\theta_1))$, $\hbar\omega^+_2=2d(k_F^+(\theta_2))$
(corresponding to the minimum and maximum of the function in $(i)$, respectively), 
and $\hbar\omega^-_1=2d(k_F^-(\theta_1))$, $\hbar\omega^-_2=2d(k_F^-(\theta_2))$
(the minimum and maximum of the function in $(ii)$); 
in region II, $\hbar\nu^-_{2,-}=2d(k_{F,-}^-(\theta_2))$,
 $\hbar\nu^-_{1,-}=2d(k_{F,-}^-(\theta_1))$ (minimum and maximum of the function in $(iii)$), 
and $\hbar\nu^-_{1,+}=2d(k_{F,+}^-(\theta_1))$,
$\hbar\nu^-_{2,+}=2d(k_{F,+}^-(\theta_2))$ (minimum and maximum of the function in $(iv)$);
in region III, only $\hbar\nu^-_{2,-}=2d(k_{F,-}^-(\theta_2))$ and  
$\hbar\nu^-_{2,+}=2d(k_{F,+}^-(\theta_2))$ remain. These energies are listed in order of increasing magnitude (Fig.~\ref{CriticalF}). 
Noting that $\hbar\nu^-_{2,-}=\hbar\omega^+_2$, $\hbar\nu^-_{1,-}=\hbar\omega^+_1$,
$\hbar\nu^-_{1,+}=\hbar\omega^-_1$, and $\hbar\nu^-_{2,+}=\hbar\omega^-_2$, the
explicit expressions for the critical frequencies are: 
\begin{eqnarray}
\hbar\omega_1^\pm(\varepsilon_F) &=&
2\Big|\sqrt{\varepsilon_R^2+2\varepsilon_F\varepsilon_R}\mp \varepsilon_R\Big|\,,\\
\hbar\omega_2^\pm(\varepsilon_F)&=&\frac{2}{1-\mathcal{D}^2}\left[2\varepsilon_R+\mathcal{D}^2\left(\varepsilon_F-\varepsilon_R \mp
\sqrt{\varepsilon_R^2+2\varepsilon_F\varepsilon_R+\varepsilon_F^2{\cal D}^2}\right)\right]^{1/2}
\!\left(\varepsilon_F+\varepsilon_R \mp
\sqrt{\varepsilon_R^2+2\varepsilon_F\varepsilon_R+\varepsilon_F^2{\cal D}^2}\right)^{1/2}\,.
\end{eqnarray}
For $\varepsilon_F< 0$, the substitution $\varepsilon_F\rightarrow|\varepsilon_F|$ is implicit. 
Note that $\hbar\omega_1^{\pm}$ is independent of the parameter $\mathcal{D}$. Fig.\,\ref{CriticalF}
displays the critical frequencies over the entire range of $\varepsilon_F$
taking ${\cal D}=0.8$. In region I, $\hbar\omega_1^+$ and $\hbar\omega_2^+$ are very
close, with $\hbar\omega_1^+\lesssim \hbar\omega_2^+$. In region II, however,
$\hbar\omega_2^+\lesssim \hbar\omega_1^+$, for the parameters used in this work.
Note also the coincidence of $\hbar\omega_1^+$ and $\hbar\omega_1^-$ at
$\varepsilon_F=-\varepsilon_R/2$.


The calculations show that
$\text{Re}\gamma^{inter}_{xx}(\omega)$ and $\text{Im}\gamma^{inter}_{xy}(\omega)$ 
are given by the principal value integrals in (\ref{ReGamma}) and (\ref{ImGamma}),
given that the integrals of the second terms vanish due to the symmetries of the Zeeman QGT (see Sec.\,\ref{SymmModel}). 
\begin{figure}
    \centering
    \includegraphics[scale=0.47]{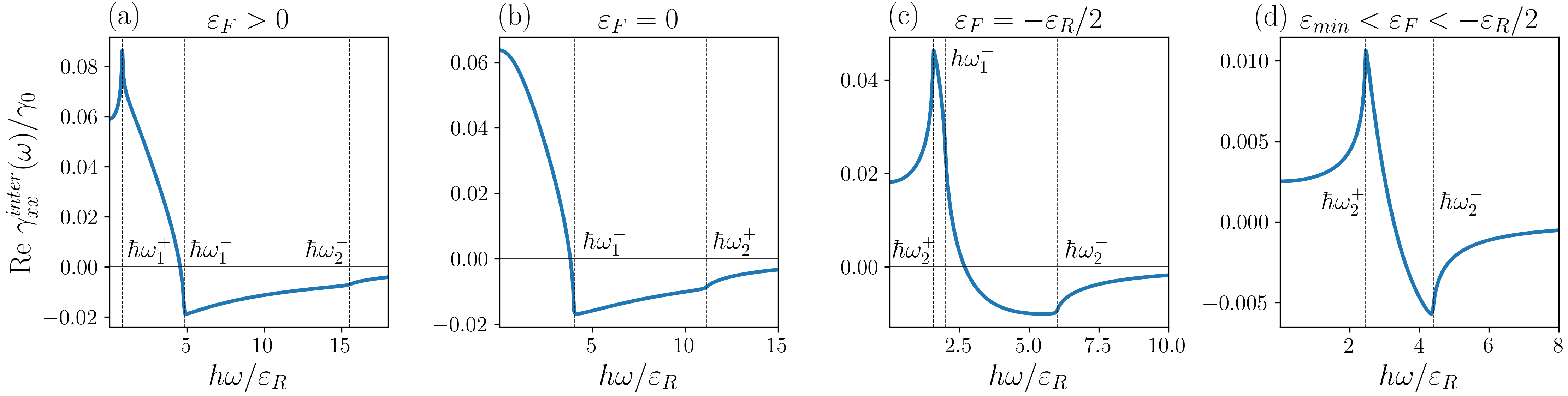}\\
    \includegraphics[scale=0.47]{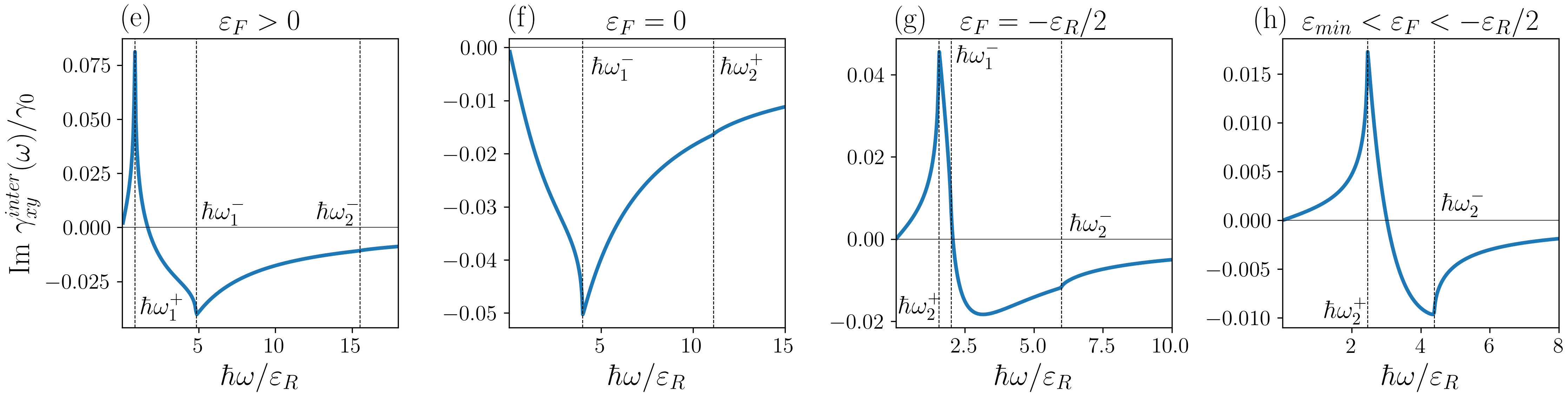}
    \caption{Reactive-like components of the spin Edelstein susceptibility tensor,
    for several values of the Fermi energy. The spectrum of $\text{Re}\gamma_{xx}^{inter}(\omega)$
    is determined by the ZQM component $\bar{g}^{xx}({\bf k})$, and that of 
    $\text{Im}\gamma_{xy}^{inter}(\omega)$ by the ZBC component $\bar{\Omega}^{yx}({\bf k})$.
    The vertical dotted lines
    indicate the position of the critical frequencies in each case. The variation in the
    number of critical frequencies as $\varepsilon_F$ moves from interval I to interval III
    reflects the topological changes of the Fermi surface. We take $\mathcal{D}=0.8$. 
}
    \label{Inter-Regammaxx_Imgammaxy}
\end{figure}
\begin{figure}[h!]
    \centering
    \includegraphics[scale=0.47]{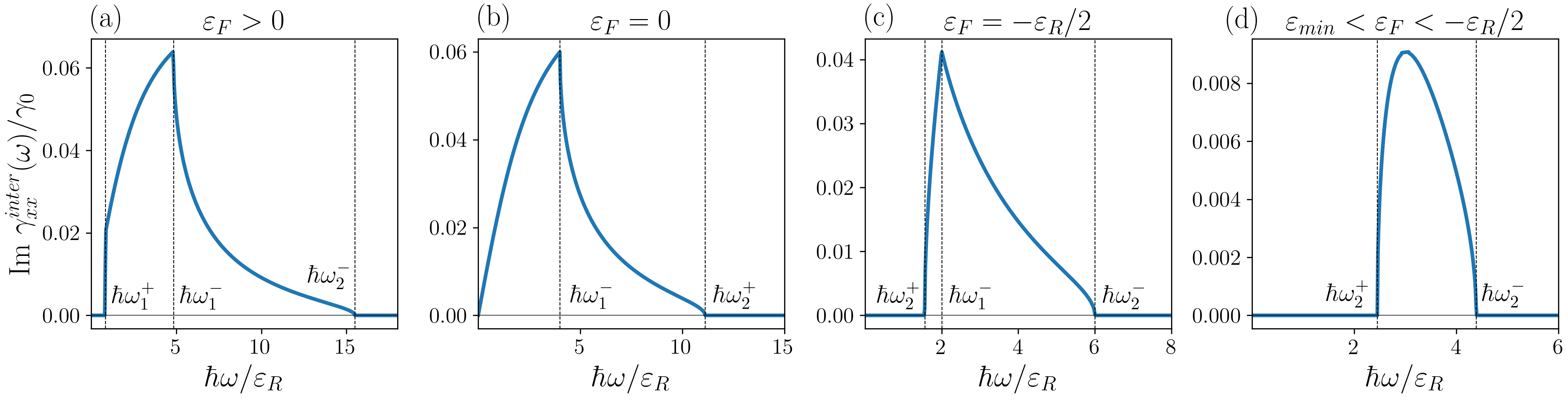}\\
    \includegraphics[scale=0.47]{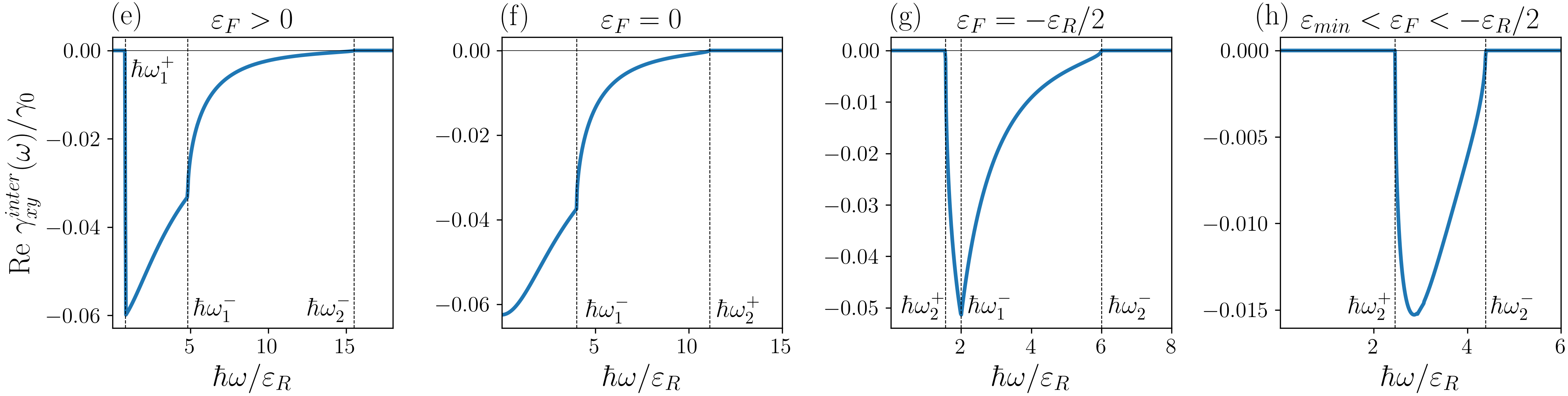}
    \caption{Absorptive-like components of the spin Edelstein susceptibility tensor, for several values of the Fermi energy. The spectrum of $\text{Im}\gamma_{xx}^{inter}(\omega)$
    is determined by the ZQM component $\bar{g}^{xx}({\bf k})$, and that of 
    $\text{Re}\gamma_{xy}^{inter}(\omega)$ by the ZBC component $\bar{\Omega}^{yx}({\bf k})$.
    The vertical dotted lines indicate the position of the critical frequencies in each case.
    At zero frequency
    $\text{Re}\gamma_{xy}^{inter}(0)=0$ except at $\varepsilon_F=0$, where
    it takes the magnitude $-\gamma_0/16$;
    in contrast, $\text{Im}\gamma_{xy}^{inter}(0)=0$ regardless of the position of 
    $\varepsilon_F$ (Fig.~\ref{Inter-Regammaxx_Imgammaxy}).
    We take $\mathcal{D}=0.8$.}
    \label{Inter-Imgammaxx_Regammaxy}
\end{figure}
On the other hand, $\text{Im}\gamma^{inter}_{xx}(\omega)$ and 
$\text{Re}\gamma^{inter}_{xy}(\omega)$ arise exclusively from the $\bar{g}\delta(\cdots)$
and $\bar{\Omega}\delta(\cdots)$ terms in
(\ref{ImGamma}) and (\ref{ReGamma}), respectively; the principal value terms integrate to zero. 
As a result, we find for the present model that $\gamma_{xx}(\omega)=\gamma^{inter}_{xx}(\omega)$ is related to interband transitions with a strength determined by the longitudinal component $\bar{g}^{xx}({\bf k})$ of the ZQM tensor, while the interband Hall component $\gamma^{inter}_{xy}(\omega)$ is determined in momentum space by the ZBC $\bar{\Omega}^{yx}({\bf k})$.
Moreover, as discussed in Sec.\,\ref{SymmModel},
the longitudinal components of the interband contribution vanish for $\mathcal{D}=0$, and become finite as long as $\alpha\mathcal{D}\neq0$, as seen by the structure of the ZQM in Table \ref{Table4}.

The spectra shown in Fig.~\ref{Inter-Regammaxx_Imgammaxy} display the `reactive' response
functions $\text{Re}\gamma^{inter}_{xx}(\omega)$ and $\text{Im}\gamma^{inter}_{xy}(\omega)$
for several positions of the Fermi level. The `absorptive' response functions 
$\text{Im}\gamma^{inter}_{xx}(\omega)$ and $\text{Re}\gamma^{inter}_{xy}(\omega)$ are shown
in Fig.~\ref{Inter-Imgammaxx_Regammaxy}.
The variation in the number of critical frequencies as the Fermi energy changes
through the three regions reflects the change in the topology of the Fermi contour. 
In region I, there are four critical frequencies.
The almost circular shape of the Fermi contour $k_F^+(\theta)$ 
(Fig.~\ref{FermiCont}(a)) makes the separation of frequencies $\omega_1^+, \omega_2^+$
very small (Fig.~\ref{CriticalF}), and the associated spectral features are barely resolved in 
Fig.~\ref{Inter-Regammaxx_Imgammaxy}(a) and Fig.~\ref{Inter-Imgammaxx_Regammaxy}(a), 
in contrast to the clearly discernible features at $\omega_1^-, \omega_2^-$.
In region II, all spectra (not shown) look very similar to those in region I.
Notably, at $\varepsilon_F=0$ (Fig.~\ref{FermiCont}(b)), the set of critical frequencies reduces to two:
$\hbar\omega_1^+=\hbar\omega_2^+=0$, $\hbar\omega_1^-=4\varepsilon_R$,
$\hbar\omega_2^-=4\varepsilon_R/(1-{\cal D}^2)$ (second column in Fig.~\ref{Inter-Regammaxx_Imgammaxy} and Fig.~\ref{Inter-Imgammaxx_Regammaxy}), 
while at the special value $\varepsilon_F=-\varepsilon_R/2$ (Fig.~\ref{FermiCont}(c)) there are three critical frequencies:
$\hbar\omega_1^{\pm}=2\varepsilon_R$ and
$\hbar\omega_2^{\pm}=\varepsilon_R(2\pm {\cal D})/(1\pm{\cal D})$
(third column in Fig.~\ref{Inter-Regammaxx_Imgammaxy} and Fig.~\ref{Inter-Imgammaxx_Regammaxy}). 
In region III, there are two spectral features associated with the two critical frequencies $\hbar\omega_2^{\pm}$, as Fig.~\ref{CriticalF} and Fig.~\ref{FermiCont}(d) illustrate.

From an experimental perspective, taking a realistic value of the Rashba parameter
$\alpha=0.2\,$eV\,\AA, ${\cal D}=0.8$, and $E\sim 1\cdots 5\,$V/nm \cite{SpinPolEng,Iurov2026}, the peak magnitude scale of Figs.\,\ref{Inter-Regammaxx_Imgammaxy} and \ref{Inter-Imgammaxx_Regammaxy}, for $\hbar\omega$ close to a critical frequency, yields an estimation of a spin magnetization density 
$\sim 0.01 \cdots 0.2\,\mu_B/$\AA$^2$. For a typical area per unit cell, this implies
magnetic moments which lie in the detectable range with current techniques.

In summary, we have found that the dynamic Edelstein response of the Rashba $d$-wave altermagnet displays a strong dependence on frequency and Fermi energy, with characteristic spectral features
in clear contrast to those of a pure Rashba coupled system.
The variation in its magnitude and sign suggests an additional method of manipulation, with longitudinal electric-field-induced spin polarization provided by the simultaneous presence of RSOI and altermagnetism.
Moreover, an experimental identification of the critical frequencies would allow
us to extract spin-orbit and altermagnetic material parameters.

\section{Spin polarization}\label{Sec4}

\begin{figure}
    \centering
    \includegraphics[width=0.7\linewidth]{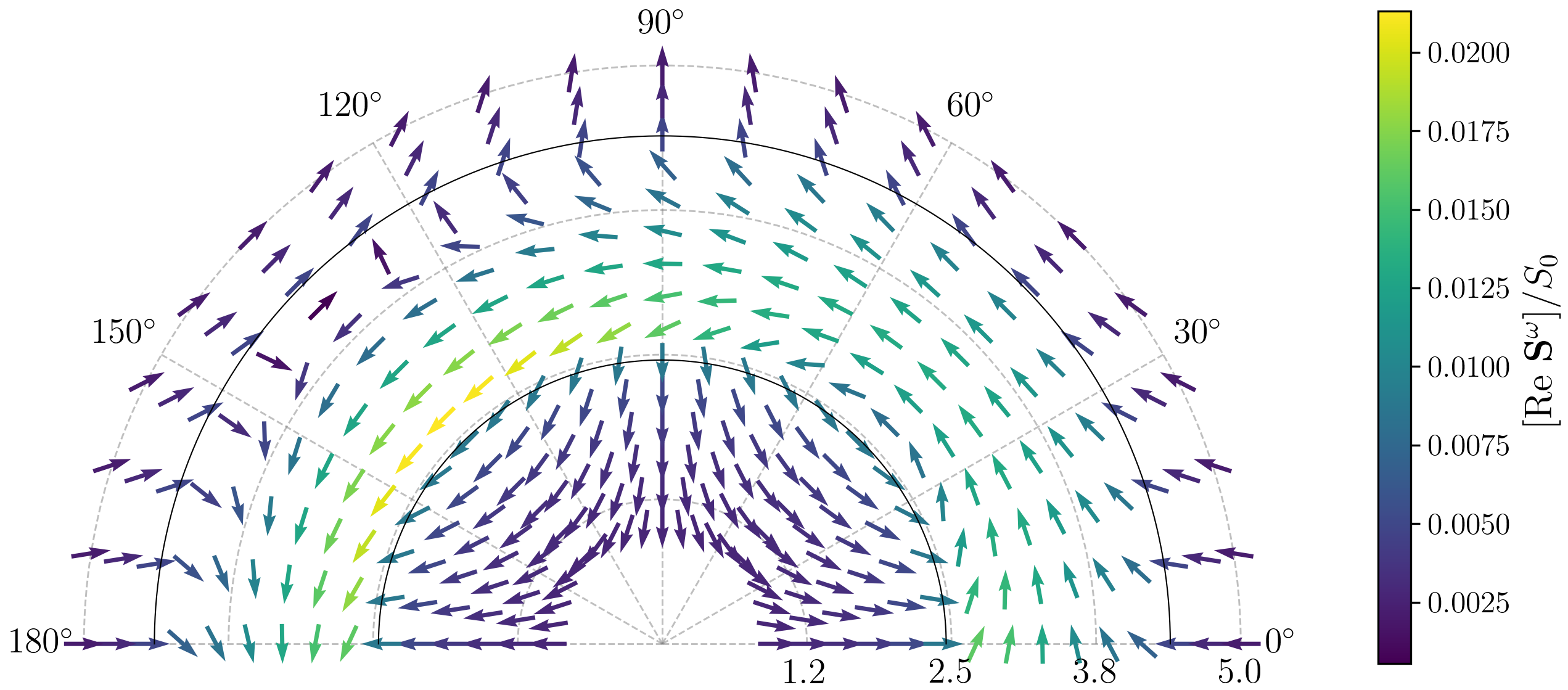}
    \caption{Real part of the complex amplitude of the induced spin polarization, 
    $[\text{Re}\,{\bf S}^{\omega}]/S_0$, for $\varepsilon_{min}<\varepsilon_F<-\varepsilon_R/2$. The arrows indicate the direction of the field and the color 
    its magnitude, according to the color bar. Given that $\text{Re}\gamma^{intra}_{xy}(\omega)=0$, only interband transitions contribute to this quantity.
    The radial coordinate corresponds to the driving frequency, while the
    angular coordinate corresponds to the orientation angle of the in-plane electric
    field. The inner (outer) black semicircle is at the critical frequencies $\hbar\omega_2^+$ 
    ($\hbar\omega_2^-$). 
    We take $\mathcal{D}=0.8$.}
    \label{Svector}
\end{figure}
The results show the following structure for the spin Edelstein susceptibility:
\begin{equation}
\gamma_{ij}(\omega)=
\begin{pmatrix}
\gamma_{xx}^{inter}(\omega) & \gamma_{xy}^{intra}(\omega)+\gamma_{xy}^{inter}(\omega) \\
-[\gamma_{xy}^{intra}(\omega)+\gamma_{xy}^{inter}(\omega)] & -\gamma_{xx}^{inter}(\omega)
\end{pmatrix}\,,
\end{equation}
$i,j=x,y$ and $\gamma_{zj}(\omega)=0$.
Writing the amplitude of the external electric field as ${\bf E}^{\omega}=E^{\omega}(\cos\varphi{\bf\hat{x}}+\sin\varphi{\bf\hat{y}})$, the amplitude of the induced spin
polarization is given by  
${\bf S}^{\omega}=\frac{\hbar}{2}[\gamma_{xx}(\omega)E^{\omega}_x+\gamma_{xy}(\omega)E_y^{\omega}]{\bf\hat{x}}+\frac{\hbar}{2}[-\gamma_{xy}(\omega)E_x^{\omega}-\gamma_{xx}(\omega)E_y^{\omega}]{\bf\hat{y}}$,
which we rewrite as
\begin{equation}
{\bf S}^{\omega}=\frac{\hbar}{2}\left[\gamma_{\parallel}(\omega,\varphi){\bf E}^{\omega}+\gamma_{\perp}(\omega,\varphi){\bf E}^{\omega}\times{\bf\hat{z}}\right],
\end{equation}
where the functions $\gamma_{\parallel}(\omega,\varphi)=\gamma^{inter}_{xx}(\omega)\cos2\varphi$ and $\gamma_{\perp}(\omega,\varphi)=\gamma^{inter}_{xx}(\omega)\sin2\varphi+\gamma_{xy}(\omega)$
characterize the anisotropy of the spin polarization with respect to directions
along and perpendicular to the applied field, respectively. The interband origin of $\gamma_{xx}$
has been emphasized. 
Thus, $\text{Re}\,{\bf S}(t)=\text{Re}({\bf S}^{\omega}e^{-i\omega t})=
[\text{Re}\,{\bf S}^{\omega}]\cos\omega t+[\text{Im}\,{\bf S}^{\omega}]\sin\omega t$, where
 the real (imaginary) part of ${\bf S}^{\omega}$ is determined by the real (imaginary)
 parts of $\gamma_{\parallel}(\omega,\varphi)$ and $\gamma_{\perp}(\omega,\varphi)$.
Note that in our clean limit ($\eta=0^+$) calculations, $\text{Re}\,{\bf S}^{\omega}$ involves $\text{Re}\,\gamma_{ij}$ components originating from interband transitions only,
because $\text{Re}\gamma^{intra}_{xy}(\omega)=0$ (see Eq.\,(\ref{IntraB2})).

To illustrate the anisotropy introduced by the frequency dependence of the
spin Edelstein susceptibility and the orientation of the electric field, we consider the
dimensionless vector field ${\bf P}(\omega,\varphi)\equiv [\text{Re}\,{\bf S}^{\omega}](\omega,\varphi)/S_0$ in Fig.~\ref{Svector}, 
for $\varepsilon_F$ lying in the interval III, close to the bottom of the $\lambda=-$ band.
We normalize by the quantity $S_0=(\hbar/2)\gamma_0E^{\omega}$.
The arrows indicate the direction of ${\bf P}$ and the color indicates its magnitude
$|{\bf P}|=[(\text{Re}\,\gamma_{\parallel}/\gamma_0)^2+(\text{Re}\,\gamma_{\perp}/\gamma_0)^2]^{1/2}$.
For example, for ${\bf E}^{\omega}=(\pm{\bf\hat{x}}+{\bf\hat{y}})/\sqrt{2}$,
${\bf P}(\omega,\pi/2\mp\pi/4)=-\text{Re}(\gamma^{inter}_{xy}\pm\gamma^{inter}_{xx})
\boldsymbol{\hat{\varphi}}/\gamma_0$, where $\boldsymbol{\hat{\varphi}}=-\sin\varphi{\bf\hat{x}}+\cos\varphi{\bf\hat{y}}$ gives the azimuthal direction in the plane of the sample. The resulting direction of this azimuthally-oriented vector field can be fixed by
tuning the frequency. If $\hbar\omega\notin [\hbar\omega_2^+,\hbar\omega_2^-]$, $\text{Re}\,\gamma_{xy}^{inter}(\omega)=0$ (Fig.~\ref{Inter-Imgammaxx_Regammaxy}(h)),
then the resulting orientation is controlled by the frequency dependence of
$\text{Re}\,\gamma_{xx}^{inter}(\omega)$ (Fig.~\ref{Inter-Regammaxx_Imgammaxy}(d)),
which is positive (negative) for $\hbar\omega$ below (above) $\omega_2^+$ ($\omega_2^-$).
For $\hbar\omega\in [\hbar\omega_2^+,\hbar\omega_2^-]$, $\text{Re}\,\gamma_{xy}^{inter}(\omega)<0$, the resulting direction of ${\bf P}(\omega,\pi/2\mp\pi/4)$ will be defined by
the balance between $\text{Re}\,\gamma^{inter}_{xy}(\omega)$ and 
$\text{Re}\,\gamma^{inter}_{xx}(\omega)$. At $\varphi=\pi/4$, there is a change of sign 
($-\boldsymbol{\hat{\varphi}}\to \boldsymbol{\hat{\varphi}}$) in ${\bf P}$ for $\omega\approx\omega_2^+$, while at $\varphi=3\pi/4$ the inversion of sign 
($\boldsymbol{\hat{\varphi}}\to -\boldsymbol{\hat{\varphi}}$) occurs
for higher frequencies, closer to $\hbar\omega_2^-$.
For ${\bf E}^{\omega}\parallel\pm{\bf\hat{x}}$, the induced vector field is
${\bf P}(\omega,\pi/2\mp\pi/2)=\pm\text{Re}(\gamma_{xx}^{inter}/\gamma_0){\bf\hat{x}}\mp\text{Re}(\gamma_{xy}^{inter}/\gamma_0){\bf\hat{y}}$ and for
 ${\bf E}^{\omega}\parallel{\bf\hat{y}}$, 
${\bf P}(\omega,\pi/2)=\text{Re}(\gamma_{xy}^{inter}/\gamma_0){\bf\hat{x}}-\text{Re}(\gamma_{xx}^{inter}/\gamma_0){\bf\hat{y}}$; the variation of the driving frequency
through the interval $[\hbar\omega_2^+,\hbar\omega_2^-]$ serves as a region of transition
to invert the sign of ${\bf P}$.
Regarding the quantity $|{\bf P}|$, the pattern of colors
in Fig.~\ref{Svector} also reveals the possibilities of manipulation introduced by the
driving frequency and the electric field orientation on such a magnitude.






\section{Conclusions}\label{Sec5}

In this work, we investigate the intraband and interband Edelstein response of a two-dimensional
$d$-wave altermagnet with Rashba spin-orbit interaction.
The presence of altermagnetism provides distinctive additional features in the band structure and associated density of states in contrast to a conventional Rashba gas. 
We find that tuning the Fermi level in the entire energy range drives  Lifshitz transitions in the Fermi contours, accompanied by a logarithmic van Hove singularity in the density of states at energy $-\varepsilon_R/2$.
Below this energy, four disconnected Fermi contours emerge before eventually collapsing into four isolated points at the band minimum.
These changes in the Fermi contour topology are reflected in the intraband Edelstein response and determine the available phase space for interband transitions.

We show how the spectrum of interband optical transitions is governed by the momentum-dependent Zeeman quantum geometric tensor. 
The symmetry properties of this tensor lead to a spin susceptibility tensor with the structure $\gamma_{xx}=-\gamma_{yy}$, $\gamma_{xy}=-\gamma_{yx}$, 
and $\gamma_{zx}=\gamma_{zy}=0$. Thus, the induced spin polarization lies strictly in the plane of the sample.
As a function of frequency, the interband spin susceptibility tensor can be divided into dissipative-type components ($\text{Im}\,\gamma_{xx}$, $\text{Re}\,\gamma_{xy}$),
characterized by a fixed overall sign, and reactive-type components  ($\text{Re}\,\gamma_{xx}$, $\text{Im}\,\gamma_{xy}$), which present a change of sign as a clear
signature. 
We find that the longitudinal interband response function $\gamma_{xx}(\omega)$ emerges from the Zeeman quantum metric, 
while the Hall interband component $\gamma_{xy}(\omega)$ arises from the Zeeman Berry curvature function. 
The altermagnetic contribution to these quantities enters as proportional to $\alpha{\cal D}$, underscoring the important role of Rashba coupling.
A particularly relevant result is the emergence of a finite longitudinal interband response only when both Rashba coupling and altermagnetism are present, in contrast to the conventional Rashba gas, whose interband response is exclusively transverse.
The response functions $\gamma^{inter}_{xx}(\omega),\gamma^{inter}_{xy}(\omega)$ develop characteristic spectral features, notably a variable number of critical frequencies
as the Fermi level changes from the positive to the negative part of the energy band structure. Such spectral behavior can be viewed as an optical fingerprint of
the topological changes of the Fermi surface. 
When combined with the electric field orientation angle, such a strong frequency dependence suggests
new possibilities for control. We demonstrate that the magnitude and direction of the spin polarization vector field can be tuned by choosing the direction and frequency
of the external electric field appropriately.

We hope the present work on the dynamical spin Edelstein effect will be useful in the optical probing and control of the spin in realistic altermagnetic materials and their potential use in optically driven spintronic applications.

\section*{Acknowledgments}\label{acknowledgements}

The authors acknowledge financial support from DGAPA-UNAM through PAPIIT project IN112125, México. M.A.M. acknowledges support from the U.S. Department of Energy, Office of Basic Energy Sciences, Materials Sciences and Engineering Division. R.C.-B. acknowledges support from the Robert and René Glidden Visiting Professorship Program at Ohio University (2025–2026), the Nanoscale Quantum Phenomena Institute (NQPI) at Ohio University, and SECIHTI (México) through the Sabbatical Leave Program (2025–2026). He also thanks Dr. Nancy Sandler and the Department of Physics and Astronomy at Ohio University for their hospitality and support during his sabbatical leave.

\bibliography{biblio.bib}
\end{document}